\documentclass[12pt,preprint]{aastex}
\usepackage{psfig}
\newcommand{\mjybm}{\mbox{mJy~beam${}^{-1}$}}
\newcommand{\sgrastar}{Sgr A$^{*}$}
\newcommand{\htwo}{\ion{H}{2}}

\begin{document}

\title{High-Resolution, Wide-Field Imaging of the Galactic Center Region at 330 MHz}

\author{Michael E.~Nord\altaffilmark{1}, T.~Joseph~W.~Lazio, Namir~E.~Kassim}
\affil{Naval Research Laboratory}
\affil{Code~7213, Naval Research Laboratory, Washington, DC
	20375-5351}
\email{Michael.Nord@nrl.navy.mil}
\email{Joseph.Lazio@nrl.navy.mil}
\email{Namir.Kassim@nrl.navy.mil}
\altaffiltext{1}{Doctoral Student, University of New Mexico}

\author{S.~D.~Hyman}
\affil{Department of Physics and Engineering, Sweet Briar College, Sweet Briar, VA 24595}
\email{shyman@sbc.edu}

\author{T.N. LaRosa}
\affil{Department of Biological and Physical Sciences, Kennesaw State University, 1000 Chastain Road, Kennesaw, GA 30144}
\email{ted@avatar.kennesaw.edu}

\author{C. Brogan}
\affil{Institute for Astronomy, 640 North A'ohoku Place, Hilo, HI 96720}
\email{cbrogan@ifa.hawaii.edu}

\and

\author{N.~Duric}
\affil{Department of Physics and Astronomy, University of New Mexico,
	800 Yale Boulevard NE, Albuquerque, NM  87131}
\email{duric@tesla.phys.unm.edu}

\begin{abstract}

We present a wide field, sub-arcminute resolution VLA image of the Galactic Center region at 330 MHz.  With a resolution of $\sim 7\arcsec \times 12\arcsec$ and an RMS noise of 1.6 \mjybm, this image represents a significant increase in resolution and sensitivity over the previously published VLA image at this frequency.  The improved sensitivity has more than tripled the census of small diameter sources in the region, has resulted in the detection of two new Non Thermal Filaments (NTFs), 18 NTF candidates, 30 pulsar candidates, reveals previously known extended sources in greater detail, and has resulted in the first detection of Sagittarius A$^*$ in this frequency range.

A version of this paper containing full resolution images may be found at http://lwa.nrl.navy.mil/nord/AAAB.pdf.

\end{abstract}

\keywords{Galaxy: center --- radio continuum: general --- techniques: interferometric}


\section{Introduction}\label{sec:intro}

At a distance of only 8~kpc, the Galactic center (GC) offers an unparalled site for examining the environment of a (moderately) active galactic nucleus.  A multi-wavelength approach is essential to understanding the diverse range of phenomena in the GC, and low-frequencies ($\nu < 1000$~MHz) provide several crucial benefits in obtaining a complete picture of the GC.  At 330 MHz, thermal sources such as classical \htwo\ regions have not yet become self-absorbed while non-thermal sources such as supernova remnants (SNRs) are typically detected easily.  Thus the interactions between these sources (e.g.~in regions of massive star formation) can be studied.  More generally, low frequency observations have intrinsically large fields of view, allowing the various components of the GC to be placed into a larger context.

The Galactic Center (GC) was first imaged at 330 MHz at high resolution in 1989 (Pedlar et al.~1989; Anantharamaiah et al.~1991).  Advances enabled by these early imaging programs include revealing the 7\arcmin\ radio halo around the Sagittarius~A region and constraining the 3-dimensional structure of the region through optical depth distributions.
However, imaging algorithms at the time were unable to compensate for the non-coplanar nature of the VLA.  Hence the full primary beam of the VLA at 330 MHz (FWHM 156\arcmin) was not correctly imaged and only the very center of the GC region was studied at high fidelity. 

More recently, exploiting a number of advances in imaging algorithms to compensate for the non-coplanar nature of the VLA, LaRosa et al.~(2000) re-imaged these data, forming a full field of view image.
This led to the discovery of many new sources, and provided an unparalleled census of both extended and small diameter, thermal and non thermal sources within 100 pc (projection) of the GC. This was afforded by significant advances in wide-field imaging algorithms, coupled with greatly increased computational power. However, even that effort fell short of utilizing the full resolving power of the VLA and the commensurate improved sensitivity it would have afforded. Since those data were presented, significant improvements in software, hardware, and computational power have continued to be realized. This motivated us to revisit the GC in order to achieve further improvements in resolution and sensitivity at 330 MHz.

In this paper we present analysis of our latest 330 MHz image generated from new A and B configuration data sets, which are appropriate for generating a map with a minimum of confusion noise and maximum sensitivity to smaller scale ($\lesssim 1\arcmin$) structure.  Consequently the entire GC region contained by the primary beam of the VLA has been imaged at the maximum possible resolution for the first time.  The image is centered on the radio-bright Sagittarius~A region and provides a resolution of $7\arcsec \times 12\arcsec$ and an RMS sensitivity of 1.6 \mjybm, an improvement by roughly a factor of 5 in both parameters over the LaRosa et al.~(2000) image. 

The improved sensitivity and resolution have led to the detection of at least two new Non Thermal Filaments (NTFs), 18 NTF candidates and 30 pulsar candidates.  It has also revealed previously known extended sources in greater detail and significantly increased the census of small diameter sources in the GC region.  In \S\ref{sec:observe} we describe the observations and in \S\ref{sec:reduction} we describe data reduction, image re-construction, and astrometry. 
In \S\ref{sec:point} and \S\ref{sec:sourcepop} we discuss small diameter sources, and in \S\ref{sec:extended} we present images of resolved sources including newly discovered NTFs and NTF candidates. Our conclusions are presented in section \S\ref{sec:conclude}.

\section{Observations}\label{sec:observe}

Two sets of observations were obtained as summarized in Table~\ref{tab:observe}.  The first was observed at 330 MHz in the A configuration of the VLA 
 in October 1996. 
Six MHz of total bandwidth centered on 332.5 MHz was split into 64 channels in order to enable radio frequency interference (RFI) excision as well as to mitigate the effects of bandwidth smearing (chromatic abberation). These data were from a series of observations designed to find candidate GC pulsars (i.e., small diameter, steep-spectrum objects; Lazio \& Cordes 2004).  The second set of observations were obtained in the A and B configurations of the VLA and were obtained between March 1998 and May 1999.  A total bandwidth of 3~MHz centered at 327.5 MHz was split into 32 channels.  
Unlike the archival data re-processed by LaRosa et al.~(2000), all these new data were obtained using all 27 antennas of the VLA.

\section{Data Reduction}\label{sec:reduction}
Data reduction and imaging at 330 MHz with the VLA utilizes procedures similar to those employed at centimeter wavelengths. Key differences are the need for more intensive data editing and the requirement to implement non-coplanar imaging of the full field of view in order to mitigate the confusion from the numerous extended and small diameter sources in the primary beam\footnote{A full description of low-frequency VLA data reduction procedures is at $\langle$URL:\hfill\linebreak http://rsd-www.nrl.navy.mil/7213/lazio/tutorial/$\rangle$.}. In general we followed reduction and imaging procedures analogous to the steps reported in LaRosa et al. (2000), although the speed and sophistication of many of the specialized algorithms have been greatly improved. 

Initial flux density and phase calibration were conducted in the standard manner, with Cygnus A used for bandpass calibration in the 1998 data and 3C286 used in the 1996 data. Flux density calibration was based on observations of 3C286, and initial phase calibration was obtained using the VLA calibrators B1830$-$360 and B1711$-$251.

\subsection{Radio Frequency Interference Excision}

A key issue for low frequency data reduction at the VLA is the impact of radio frequency interference (RFI).  Some sources of interference, such as lightning and solar-related activity, are normally broad-band, and require those time periods to be completely excised from the data.  However, RFI at 330 MHz is mostly narrow-band. Algorithms exist that attempt to automate the removal of only those channels with interference. 
We elected to inspect the data and remove RFI manually because 
in our experiences with automated RFI excision, either available algorithms removed too much good data, or failed to excise sufficient RFI, particularly at low-levels.

RFI excision was based on the following criteria - first, visibilities with excessive amplitudes (e.g., $> 100\sigma$) were flagged.  Then the visibility data amplitudes were scrutinized in both Stokes I and V.  Stokes V is particularly useful in locating RFI as there should be very little circular polarization at these frequencies\footnote{The radio source associated with the massive black hole in the center of our galaxy, \sgrastar, is slightly circularly polarized at higher frequencies (Bower, Falcke, \& Backer 1999).  However, the flux density of this source is very low ($<$0.1\%) compared to the total flux density in the field.} while RFI is often highly circularly polarized. Baselines and time ranges that showed excessive deviation from surrounding data were flagged.  An additional means by which RFI was localized was the identification of systemic ripples in the image. Determining the spatial frequency of these ripples allowed the offending baseline and time range to be located and removed from the visibility ($u$-$v$) dataset.

After RFI excision, the spectral line data were smoothed by a factor of two in order to lower the computational cost of imaging.  
As sensitivity declines steeply near the edge of the bandwidth, the end channels were omitted.  The resulting data set had a bandwidth of 2.34 MHz, 12 channels with $0.195$ MHz each.

\subsection{Wide-Field Imaging \& Self-Calibration}

An additional complication for low frequency imaging is that the combination of the large field of view (FWHM 156\arcmin\ at 330 MHZ), high angular resolution, and non-coplanar nature of the VLA necessitates specialized imaging algorithms to avoid
image distortion.   We employed the polyhedron algorithm of \cite{cp92}, in which the sky is approximated by many two-dimensional facets. We chose our facets to be $\sim 30\arcmin$ in size.  This choice was driven by the degree of non-coplanar image distortion deemed acceptable at facet edges.  The algorithm shifts the phase center to the center of an individual facet and then grids the $u$-$v$ channel data before it is imaged.  Iterating over many facets allows the entire primary beam to be imaged with minimal non-coplanar effects, at the minimal bandwidth smearing of the individual channels, and at the sensitivity of the full bandwidth.

Below $\sim$ 1 GHz, atmospheric phase errors for interferometers are dominated by the ionosphere.  In order to remove ionospheric phase errors, an imaging/self-calibration (Cornwell \& Formalont 1999) loop is used.  For each data set, several iterations of self-calibration were used to improve the dynamic range.  A phase self calibration interval of 2 minutes was used as this is generally short enough to track ionospheric changes and long enough to provide a sufficient signal-to-noise ratio.  Amplitude self-calibration was used only after many iterations of phase-only imaging/self-calibration loops, and utilized larger solution intervals, as described below.

Current angle-invariant implementations of self-calibration solve for one phase and/or amplitude per antenna per time interval.  For this reason, only ionospheric fluctuations with isoplanatic patch sizes on the sky large compared to the field of view can be properly removed.  Forms of angle variant self-calibration are needed to compensate for non-isoplanatic effects, especially at the lower VLA frequency of 74 MHz where those effects become severe.

The state of ionospheric weather during an observation has a strong bearing on data quality.
We were fortunate to have a very calm ionosphere 
during both A configuration observations.  In our B configuration observation, the ionosphere was less calm and data from a small number of relatively longer $u$-$v$ baselines had to be flagged for the first two hours of the observation.  However those data were compensated for by high quality A configuration data covering regions of the $u$-$v$ plane lost to the turbulent ionospheric conditions early in the B configuration.  

For the special case of the GC, most ($> 90$\%) of the flux density in the primary beam lies within the central facet containing Sgr~A. Until properly deconvolved, artifacts from Sgr~A dominate all other sources of error in the image. At the early stages of imaging, calibration and ionospheric phase errors compound this confusion problem.  Therefore, in the first imaging iteration, only the central facet containing Sgr~A was imaged.    
However, much of the emission in this field is diffuse, and standard deconvolution, which assumes point sources on an empty background, will not deconvolve this diffuse emission effectively.  Hence SDI \citep{sdi} deconvolution in AIPS was used.  SDI clean more effectively cleans diffuse emission by selecting and deconvolving all pixels above a certain intensity in an image instead of iteratively deconvolving a few bright pixels.  However, we found that starting with SDI clean resulted in the removal of too much emission from the central bright region, causing deconvolution errors in each successive major cycle.  Therefore, deconvolution was started with standard Cotton/Schwab (SGI) clean, and switched to SDI after the first major cycle. Gradually the number of facets was expanded so that successive loops of phase self-calibration and imaging encompassed the full field of view.  Once the number of facets was been expanded to include the entire field, a final amplitude and phase self-calibration with a long ($\sim$ one hour) solution interval was performed to correct for any systematic gain offsets between antennas.

\subsection{Multi-Configuration Data Synthesis}

\begin{figure}[p]
\vspace*{-0.75cm}
\begin{center}
\mbox{\psfig{file=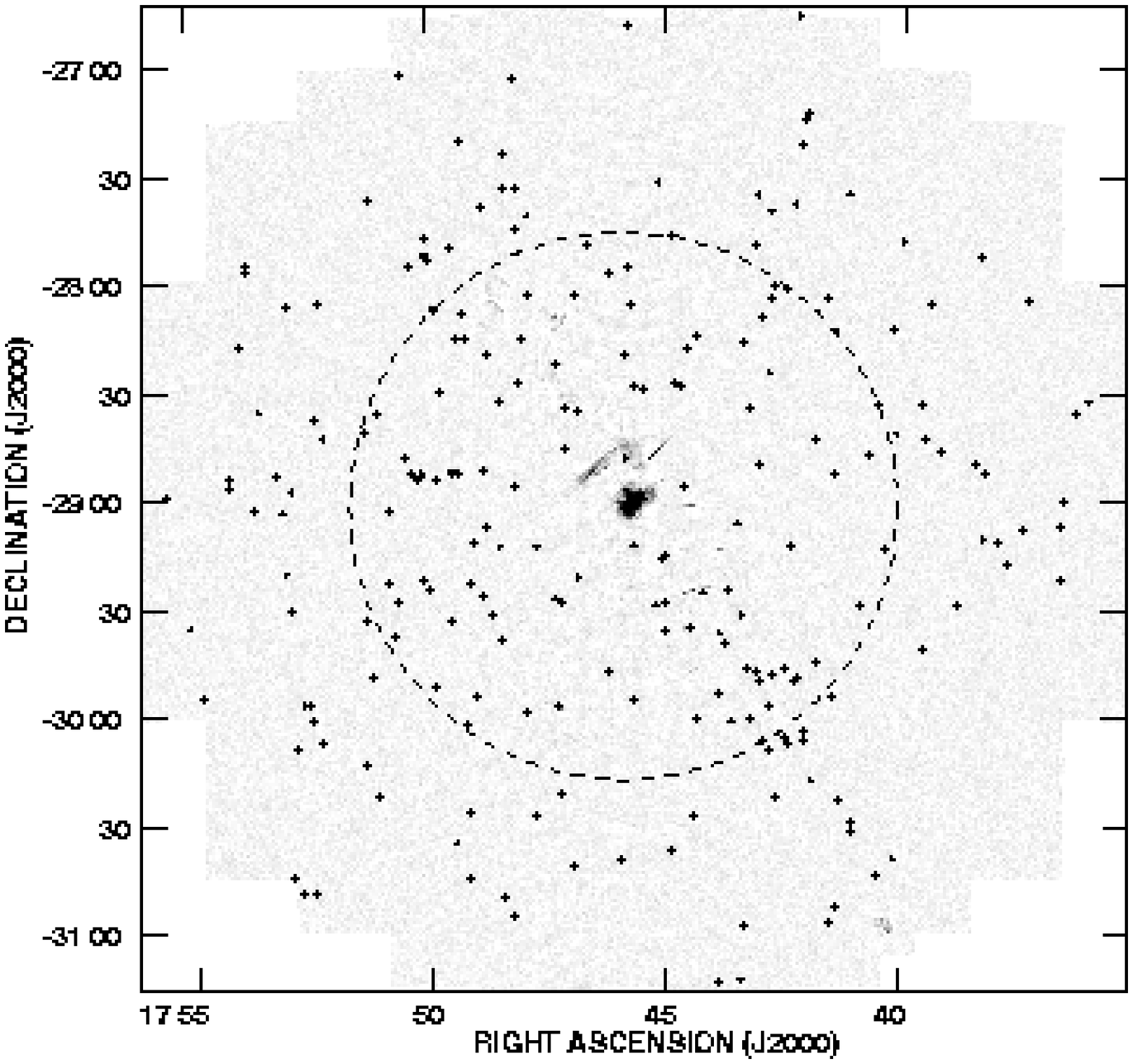,angle=0,width=1.0\textwidth,silent=}}
\end{center}
\vspace*{-1cm}
\caption[]{330 MHz A$+$B-configuration image of the Galactic Center region.  Primary beam correction has not been applied.  The dashed circle represents the half power point of the primary beam (FWHM $\sim 156\arcmin$).  The synthesized beam is $12\arcsec \times 7\arcsec$, and the RMS noise 
is 1.6~\mjybm.  The gray scale is linear between~$-2$ and 50 \mjybm.  The scalloping of the image around the edges delimits the region imaged.  Crosses indicate small diameter source locations from Table~\ref{tab:catalog}.  
}
\label{fig:gc}
\end{figure}

\begin{figure}[p]
\vspace*{-0.75cm}
\begin{center}
\mbox{\psfig{file=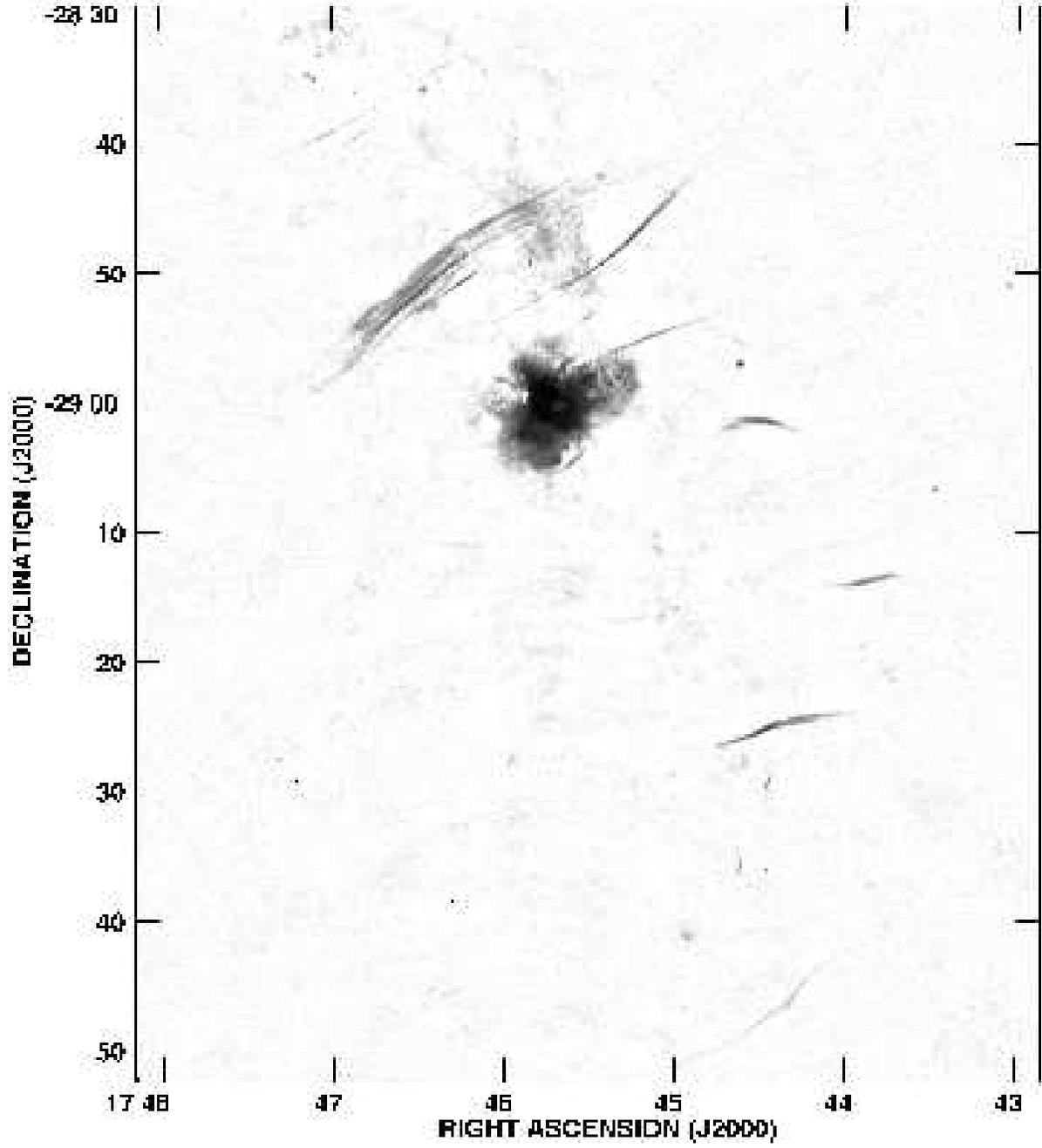,angle=0,width=1.0\textwidth,silent=}}
\end{center}
\vspace*{-1cm}
\caption[]{The inner $\sim 1.0\arcdeg \times 1.2\arcdeg$ of Figure \ref{fig:gc}.  This image was generated using a non-linear transfer function in order to show the detail in the Sgr A region and the fainter NTFs and NTF candidates.}
\label{fig:gc_center}
\end{figure}

The data from each of the three epochs were reduced separately following the procedures outlined above.  Once reasonably high dynamic range images could be produced from all three data sets, intensities of small diameter sources were checked for internal consistency.  The 1996 A configuration image was found to have small diameter source intensities which were systematically low by a factor of roughly 20\%, for reasons we could not determine.  For this reason, as well as to bring all data onto a common amplitude scale, the datasets were self-calibrated one final time.  The concatenated $u$-$v$ dataset was amplitude and phase self-calibrated with the 1998 B configuration image as the model. The self-calibration was done using a time interval of 12 hours, longer than the time of any of the individual observations.  This corrected for any systemic gain or position offsets between the datasets.  The B configuration model was chosen to anchor this alignment because use of an A configuration model would bias the flux densities to be too low.  While this technique aligned the flux density scales of the three datasets, absolute flux density calibration remains unknown at about the 5\% level (Baars et al.~1977).  After this last self-calibration, the combined data were imaged a final time, producing the final facets.  For the final image, all facets were interpolated onto one large grid, resulting in a single image with a resolution of $\sim 7\arcsec\ \times 12\arcsec$ and an RMS noise of 1.6 \mjybm.
Figure~\ref{fig:gc} shows the final image, containing over   a third of a billion $1\arcsec \times 1\arcsec$ pixels and Figure~\ref{fig:gc_center} shows the central $\sim 1.2\arcdeg \times 1.0\arcdeg$ of the field.  The total deconvolved flux density from the combined data set was 326 Jy.

\subsection{Astrometry}\label{sec:astrometry}

\begin{figure}[p]
\vspace*{-0.75cm}
\begin{center}
\mbox{\psfig{file=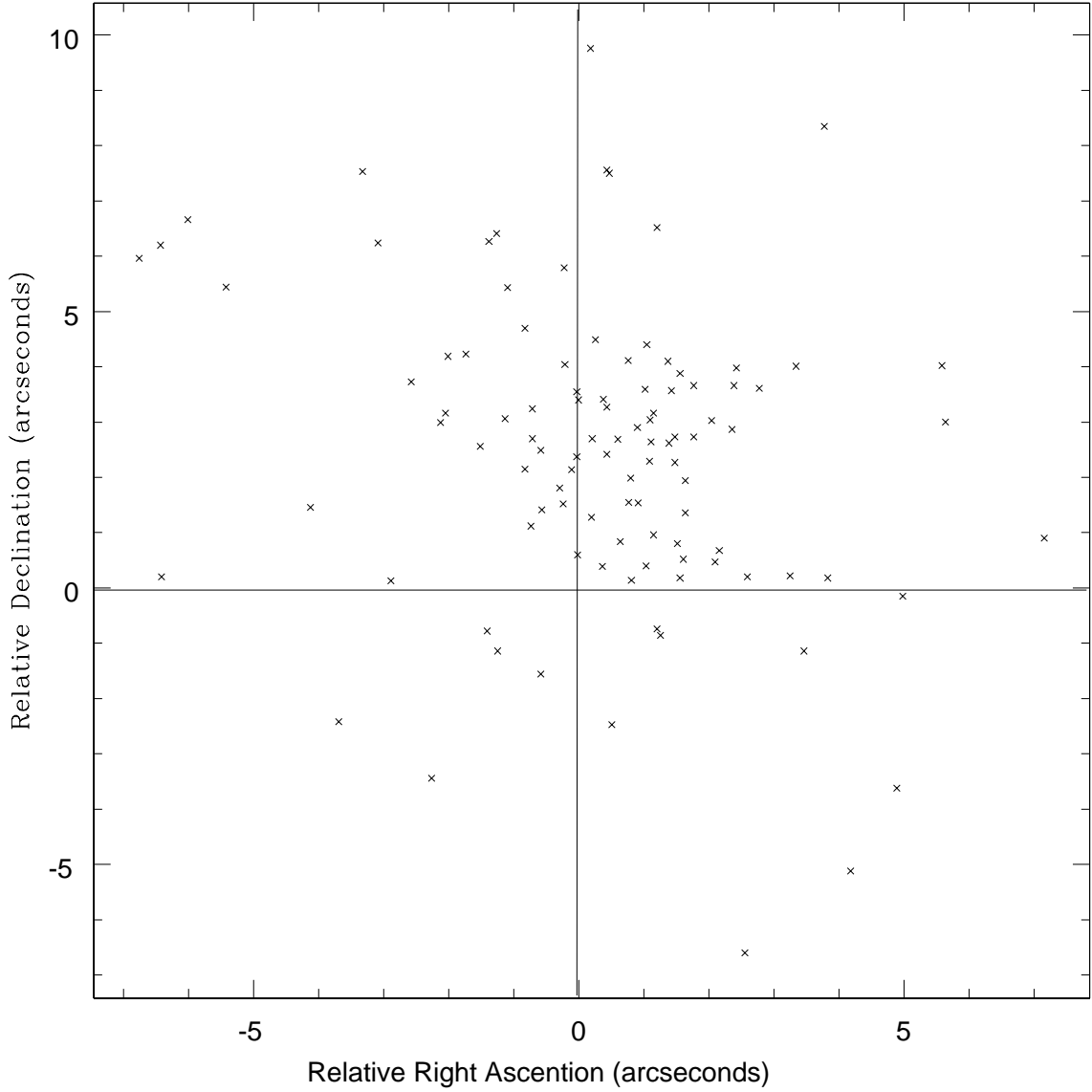,angle=0,width=1.0\textwidth,silent=}}
\end{center}
\vspace*{-1cm}
\caption[]{The position offset between the NVSS positions and the nominal positions for 103 small diameter sources common to both surveys.  The astrometric correction applied in Section \ref{sec:astrometry} was derived from these offsets.

}
\label{fig:scatter}
\end{figure}

Absolute position determination for low frequency images inevitably relies on tying their coordinates to a grid of sources whose positions are determined from higher frequency maps.  The low frequency data alone are incapable of providing good astrometry for two key reasons.  First, self-calibration inherently returns an improved visibility data set whose position is arbitrarily tied to the position of an imperfect starting model.  Secondly, even prior to self-calibration the large scale component of the ionosphere introduces an arbitrary phase shift on both target field and phase calibrator observations.  Fortunately, as described by Erickson~(1984), this second effect manifests itself mainly as a global position shift, and to first order does not distort the brightness distribution within the image.  Hence to
correct for these positional inaccuracies, small diameter sources extracted from the image (\S\ref{sec:point}) were registered against the NRAO VLA Sky Survey (NVSS) 1.4 GHz survey \citep{NVSS}.  Figure \ref{fig:scatter}~shows the relative positions of the 103 matching sources.  The mean of this distribution is offset from zero by 0.37\arcsec\ in Right Ascension and 2.4\arcsec\ in Declination.  All small diameter source positions were adjusted to account for these offsets.  We define the root mean square deviation from the mean, 2.1\arcsec, as the positional accuracy of the compact sources.


\section{Small Diameter Sources}\label{sec:point}

Locating and cataloging small diameter (less than $\sim$two beam widths) sources in the GC region is challenging.  Regions of extended emission can confuse automated small diameter source detection algorithms, yet detection by eye can bias against finding weak sources.  In this data set, we have the advantage that a great deal of the extended emission in the region has been resolved out, but enough emission still exists in supernova remnants, non-thermal filaments and extended \ion{H}{2}\ regions to confuse automated searches.  For this reason, we used a hybrid small diameter source search method in which regions of extended emission were excluded from automated small diameter source searches.  These regions included the Sgr A region and the region to the northeast along the Galactic plane extending out to the Sgr D \ion{H}{2}\ region.  To the south, the non-thermal filament Sgr C and the "Tornado" supernova remnant were also removed.  From the remaining region, an automated small diameter source search algorithm\footnote{AIPS task SAD.} was used to locate sources with a signal to noise threshold exceeding $5\sigma$.  Searches by eye were then used in areas that had been removed.  Due to confusing flux density, small diameter source detection in these areas can not be considered complete.  Finally, all sources were examined by eye to exclude genuinely extended sources, sidelobe artifacts, or similarly mis-identified small diameter sources.  In total, 241 small diameter sources were identified in this manner, more than tripling the number of small diameter sources detected in LaRosa et al.~(2000).  Figure~\ref{fig:gc} shows the locations of the Galactic center P-band survey (GCPS) small diameter sources.

\begin{figure}[p]
\vspace*{-0.75cm}
\begin{center}
\mbox{\psfig{file=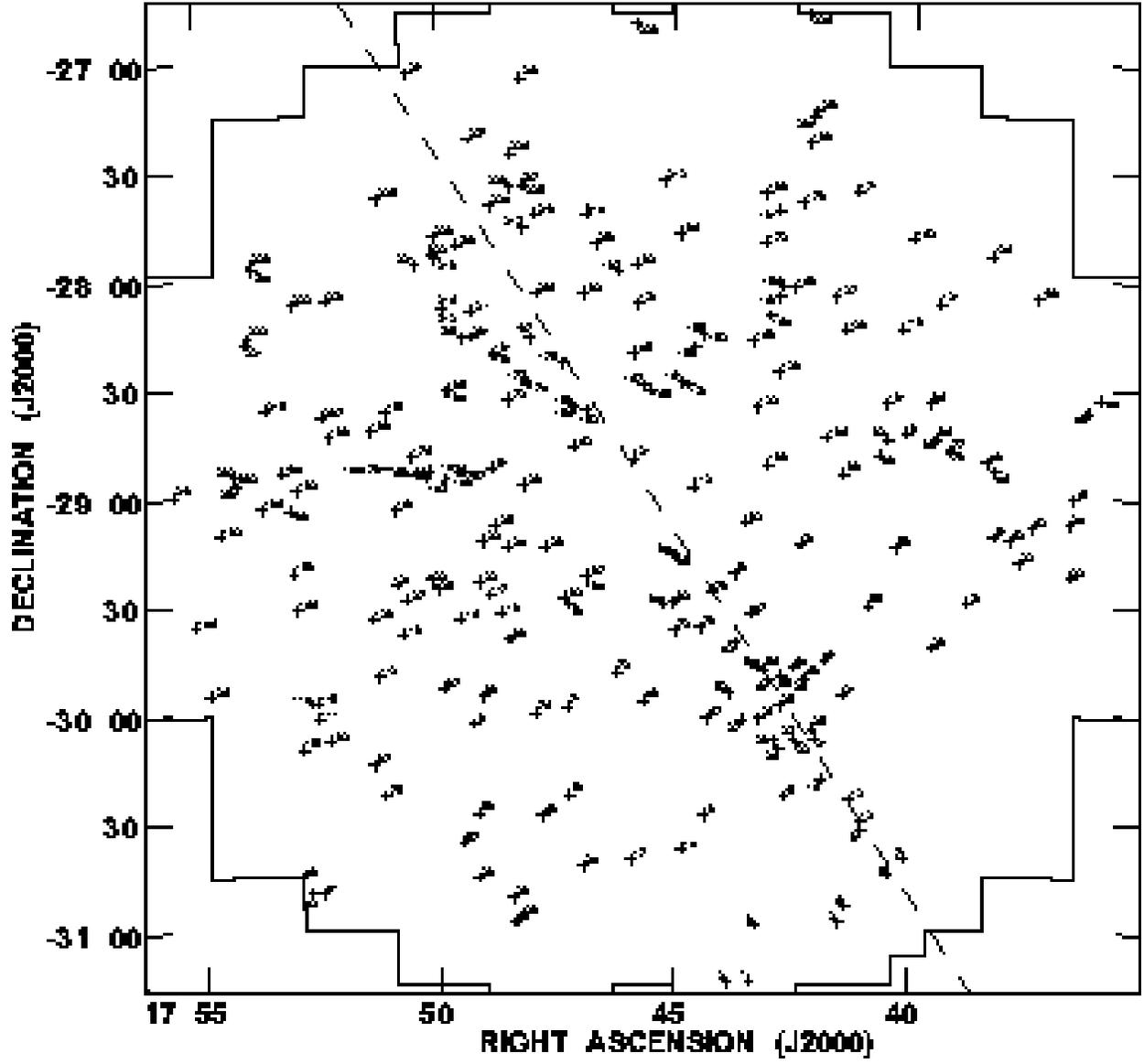,angle=0,width=1.0\textwidth,silent=}}
\end{center}
\vspace*{-1cm}
\caption[]{Map for locating sources in Table~\ref{tab:catalog}.  The dashed line represents $b=0$.}
\label{fig:finder}
\end{figure}

Once the small diameter sources had been found, two dimensional Gaussians were fit to the  sources in order to solve for positions, intensities, flux densities, and deconvolved sizes.  The distance of each source from the phase center was computed, and the resulting primary beam correction was applied.  
It should be noted that the primary beam correction is a modeled function and therefore  
flux densities of sources beyond the half power point of the primary beam ($\sim$ 80\arcmin) should be considered uncertain.  Furthermore, there are many sources in the GC region which are extended at this resolution, but are still detected by the search routine.  For this reason, we include the average of the major and minor axes of the Gaussian fit to each source.  In the cases where this value is significantly greater than the average beam size (9.75\arcsec), the source may be partially resolved, and the flux density measurement is therefore only a lower limit.  Details are given in Table \ref{tab:catalog}.  Column 1 numbers the sources, column 2 identifies sources using their Galactic coordinates, columns 3 and 4 give source RA and DEC, column 5 gives maximum intensity, column 6 gives the RMS of the image in the region local to the source, column 7 gives flux density, column 8 gives the arithmatic mean of the deconvolved major and minor axes, column 9 gives the offset of the source from the phase center, and column 10 contains information pertaining to source matches from the SIMBAD database.  Figure~\ref{fig:finder} displays the locations of sources in Table~\ref{tab:catalog}.

In order to obtain spectral information, we compared our catalog (Table \ref{tab:catalog}) against three catalogs at higher frequencies.  Table~\ref{tab:crossid} details sources having counterparts in these surveys.  Sources were considered a match if their stated location matched ours to within 15\arcsec.  At a frequency of 1.4 GHz, the GPSR (Zoonematkermani et al.~1990 \& Helfand et al.~1992) and 2LC~(Lazio \& Cordes 2004) surveys were used.  The GPSR matches our observations well in survey area and in resolution.  The 2LC has a much smaller beam size and limited coverage (inner $\sim1\arcdeg$), but was useful in resolving sources separated by less than a beam.  At 5 GHz, the companion survey to the GPSR, the GPSR5 (Becker et al.~1994) was used as it matches our resolution and coverage as well. 

Of particular note 
are two transient sources (GCRT J1746$-$2757 and XTE J1748$-$288) both described in Hyman et al.~(2002).  The former 
was discovered with this data and is undetected in the X-ray.  The latter,  
an X-ray transient, was first detected at higher radio frequencies by Hjellming
et al.~(1998).  We are presently monitoring the GC at 330 MHz several times
a year in order to constrain the frequency and magnitude of Galactic
center transients (Hyman et al.~2003).

\section{Assessing Small Diameter Source Populations}\label{sec:sourcepop}
\subsection{Small Diameter Source Density}

\begin{figure}[p]
\vspace*{-0.75cm}
\begin{center}
\mbox{\psfig{file=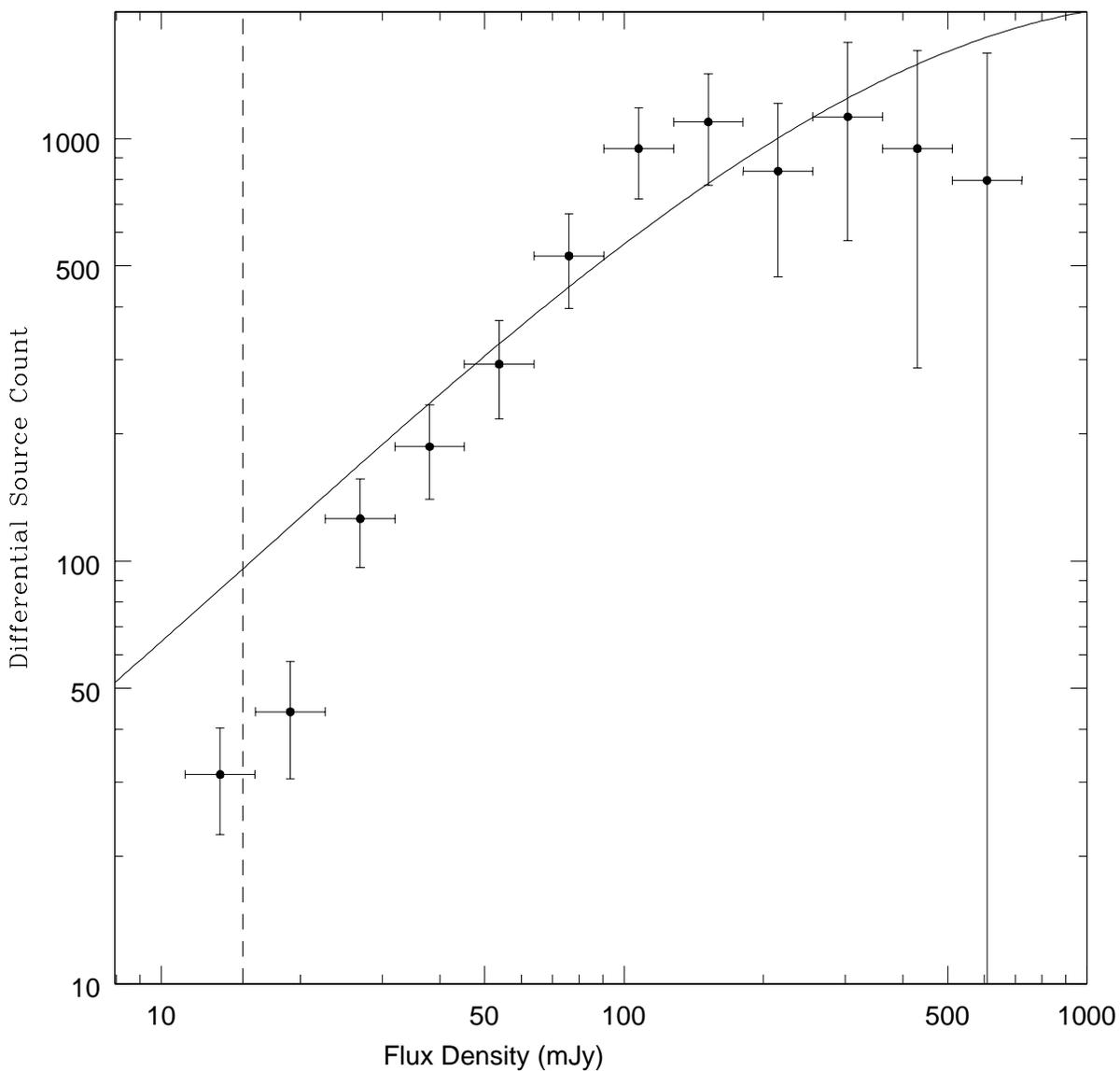,angle=0,width=1.0\textwidth,silent=}}
\end{center}
\vspace*{-1cm}
\caption[]{Euclidean normalized differential small diameter source counts for the inner 1.3\arcdeg\ of the Galactic Center image.  On the ordinate is plotted S$^{5/2} \times \frac{dN}{dS}$ in units of Jy$^{3/2}$~ster$^{-1}$.  The dashed line denotes the theoretical completeness limit, and the solid line shows the source counts from a deep WSRT survey (Wieringa 1991).  Note the increasing difference between the WSRT observed source counts and the GC source counts with decreasing flux density. }
\label{fig:dnds}
\end{figure}

As the Galactic center is one of the most densely populated regions of the sky, we expect the source density to be greater than in other regions of the sky.  To test this hypothesis, source counts from the deep WSRT (Wierenga~1991) and Cohen et al.~(2003) surveys, both at 330 MHz, were examined.  Cohen et al.~imaged a region far from the Galactic plane with the VLA with sensitivity and beam size similar to our GC image.  Within the half power point of the primary beam, and correcting for slightly greater sensitivity in the Cohen et al.~image, 209 small diameter sources were detected in Cohen et al.~versus 123 in our Galactic center image.  Wierenga (1991) used the Westerbork Synthesis array telescope to survey a large region of the sky ($\sim$ 90 square degrees), and fit a differential source count $(\frac{dN}{dS})$ model to the data. Figure~\ref{fig:dnds} shows the euclidian normalized differential source counts for the Galactic Center image with the $(\frac{dN}{dS})$ model from the deep WSRT survey (Wierenga~1991) superimposed.  The number of sources expected under this model was obtained by the numerical integration of the following:

\begin{displaymath} 
N = \int\limits_{\theta_{\mathrm{min}}}^{\theta_{\mathrm{max}}} 2\pi d\theta \int\limits_{S_{\mathrm{min}}}^{\infty} \frac {dN}{dS} dS
\end{displaymath}

Where $\theta_{\mathrm{min}}$ is the radial distance from the phase center at which the integration is started, $\theta_{\mathrm{max}}$ is the radial distance at which the integration is stopped, and $S_{\mathrm{min}}$, is the minimum detectable intensity at $\theta_{\mathrm{max}}$.  For our purposes, $\theta_{\mathrm{min}} = 10\arcmin $ in order to exclude the central Sgr A supernova remnant, $\theta_{\mathrm{max}}$ is the full width at half maximum (FWHM) of the primary beam, and $S_{\mathrm{min}}$ is 15 mJy, the $5\sigma$~detection limit at the FWHM of the primary beam.  With these values, Wieringa's source count model predicts 194 sources out to the survey limit of Cohen et al.~($1.3\arcdeg\ $radius).  The observed value of Cohen et al.~(2003) and the expected value of Wieringa (1991) agree to within 7\%.  However, the Galactic Center region's 123 sources represent an underdensity of $\sim$ 40\%.  This is at least partly explained by the presence of bright extended sources such as Sgr A, Sgr B, and Sgr C, as detection of small diameter sources that lie behind them is not possible.  However, these sources cover no more than 5\% of the region.  To explain this underdensity we hypothesize that the free electron scattering screen of Cordes and Lazio (2004) is sufficiently strong in the Galactic Center region that sources of lower intrinsic intensity are being scattered to such an extent that their surface brightness falls below the detection limit of the survey.  The scattering model and the ramifications of this observation are covered in detail in section \ref{sec:scattering}.

\subsection{Small Diameter Source Classification}\label{subsec:psourceclass}

At high Galactic latitudes the field of view at 330 MHz is dominated
by an extragalactic source population, typically radio galaxies (e.g. Cohen et al.~2003).
Given that the line of sight for this observation passes through the maximum extent of the
Galactic disk, we also expect a contribution from a Galactic source
population(s).  In this section we assess the extent to which we have
detected both a Galactic and extragalactic population of sources and in particular seek to classify the underlying nature of the small diameter component.

\begin{figure}[p]
\vspace*{-0.75cm}
\begin{center}
\mbox{\psfig{file=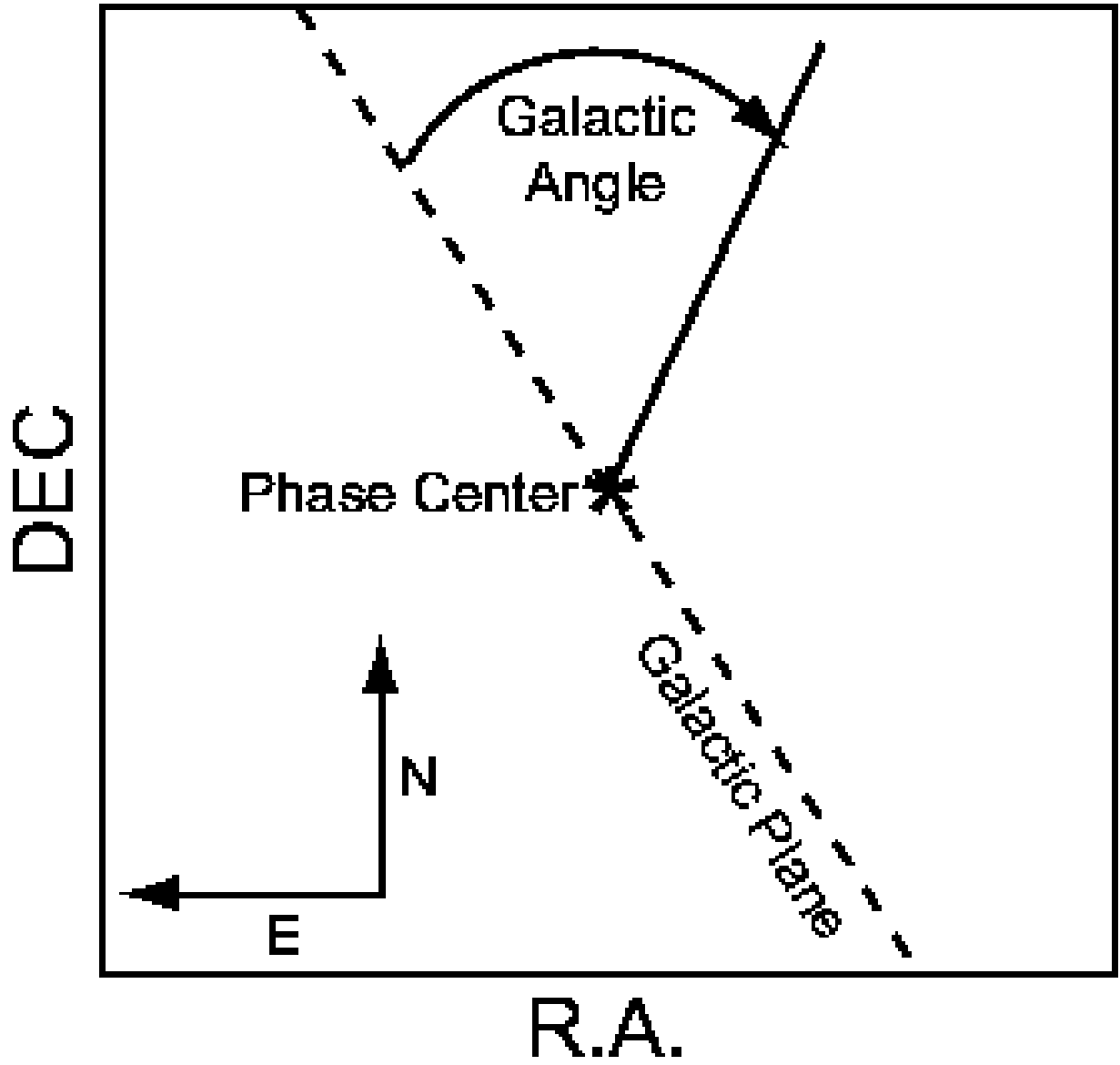,angle=0,width=1.0\textwidth,silent=}}
\end{center}
\vspace*{-1cm}
\caption[]{Schematic of the Galactic Angle coordinate system discussed in Section~\ref{subsec:psourceclass}.}
\label{fig:galangle}
\end{figure}

To determine if any Galactic population is present, the clustering of sources near the Galactic plane was examined.  Figure~\ref{fig:gc}, shows the distribution of small diameter sources.  The sources appear to be concentrated along the Galactic plane.  In order to test this observation statistically, all small diameter sources were placed into a radial coordinate frame.  The observation (phase) center is the center of the frame, the parallel to the Galactic plane toward the north is defined as zero degrees, and westward is defined as positive Galactic angle.  A schematic of this coordinate system is shown in Figure~\ref{fig:galangle}.  This coordinate system was chosen because any given angle range $d\phi$ will have equal area, $r^2 d\phi$,  where $r$ is the radius of the imaged area.  Furthermore because the sensitivity of the VLA primary beam falls off radially from the phase center any given $r^2 d\phi$ also has an equal sensitivity.  

\begin{figure}[p]
\vspace*{-0.75cm}
\begin{center}
\mbox{\psfig{file=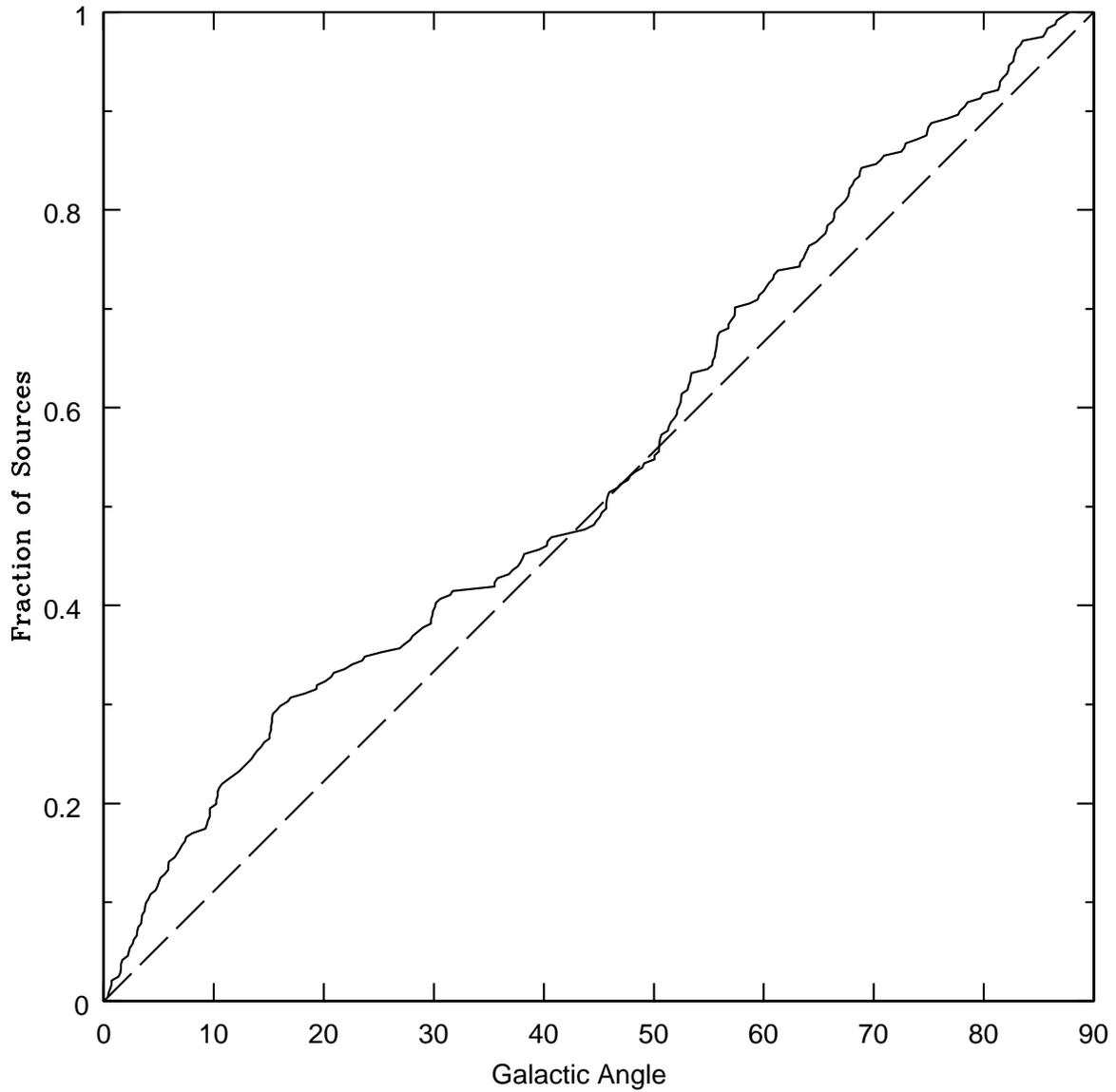,angle=0,width=1.0\textwidth,silent=}}
\end{center}
\vspace*{-1cm}
\caption[]{Graphical representation of the unbinned Kolomogorov-Smirnov test.  The solid line is the observed distribution of Galactic angle for all sources in Table~\ref{tab:catalog}, the dashed line shows the expected distribution under the null hypothesis that the sources are distributed randomly.  The observed distribution rises steeply at low Galactic angles, demonstrating that the small diameter sources tend to cluster near the plane.}
\label{fig:kolo}
\end{figure}

We assume that any Galactic or extra-galactic population would be symmetric about the Galactic plane.  Furthermore, both populations are assumed to be symmetric about the perpendicular to the Galactic plane passing through the Galactic center.  The small diameter sources were therefore reflected along these lines to place all sources in one quadrant.  In this coordinate system, sources near the Galactic plane will have small Galactic angles $(\leq 35\arcdeg)$ and sources further from the plane will have larger Galactic angles $(\geq 55\arcdeg)$.  An unbinned Kolmogorov-Smirnov (KS) test was performed on these data against the null hypothesis that the sources are randomly distributed.  The KS test, shown graphically in Figure~\ref{fig:kolo}, excludes the null hypothesis at a confidence level of 99.8\% ($3.6 \sigma$).  Furthermore, the percentage of sources rises steeply at low Galactic angle, indicating an overdensity near the Galactic plane.  The small diameter source catalog must therefore contain a component of Galactic sources, as a purely extra-galactic sample would not show clustering along the Galactic plane and indeed might be expected to show an anti-correlation due to increased scattering (Lazio \& Cordes 1998a,b) along the plane.

\begin{figure}[p]
\vspace*{-0.75cm}
\begin{center}
\mbox{\psfig{file=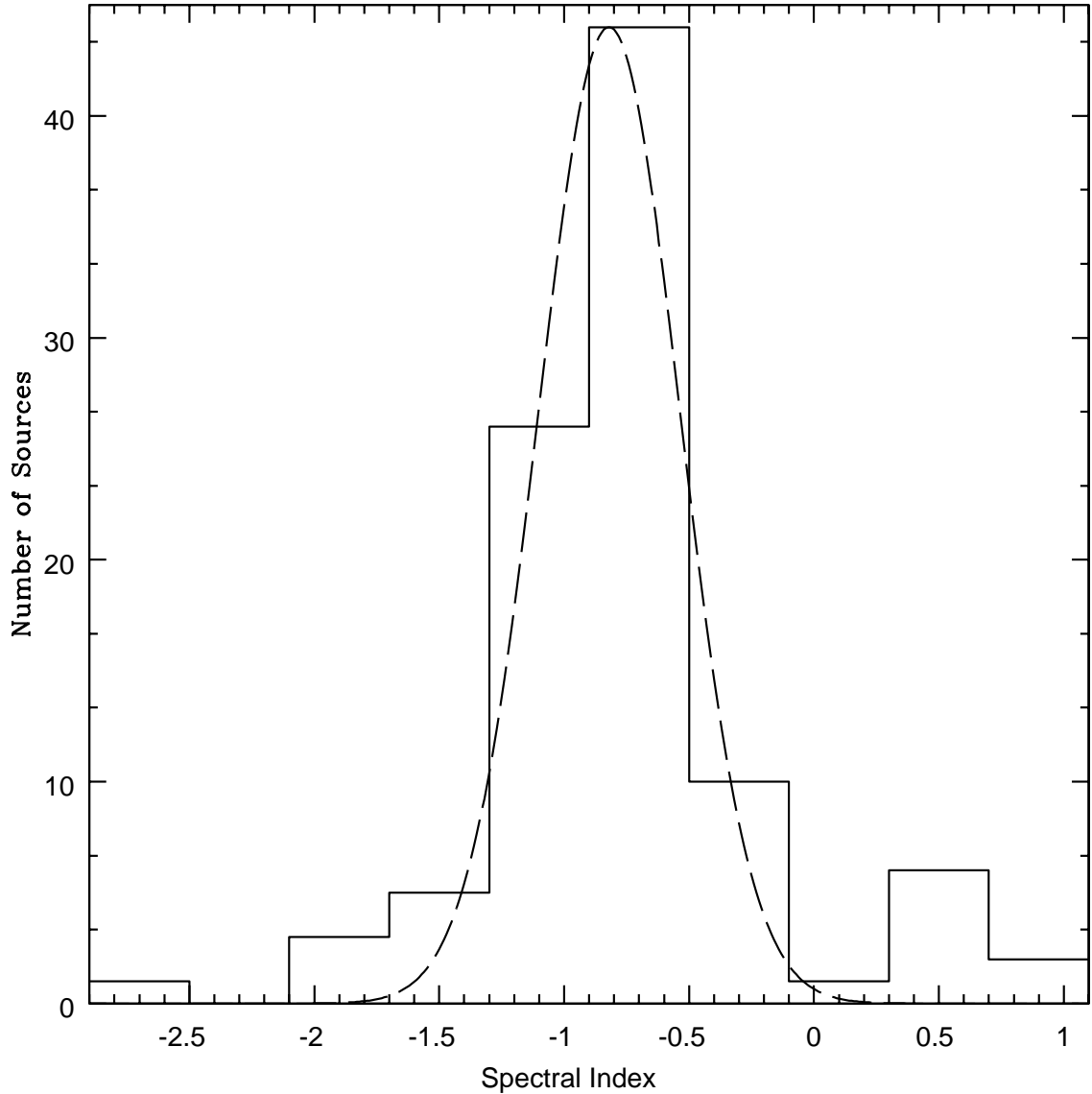,angle=0,width=1.0\textwidth,silent=}}
\end{center}
\vspace*{-1cm}
\caption[]{$\alpha_{1.4}^{0.33}$ spectral index ($S \sim \nu^{\alpha}$) histogram of the 98 sources detected at 330 MHz and in the GPSR 20 cm survey (Zoonematkermani et al.~1990 \& Helfand et al.~1992).  The Gaussian represents the distribution expected if the sources were purely extra-galactic (De Brueck et al.~2000).}
\label{fig:alphahist}
\end{figure}

A means of understanding the nature of the small diameter sources is via their spectral index ($S\propto\nu^{\alpha}$) for cases in which a higher frequency detection exists.  The sources from Table~\ref{tab:crossid} with 1.4 GHz GPSR (Zoonematkermani et al.~1990 \& Helfand et al.~1992) matches were examined as this survey has similar sensitivity and resolution to our image.  For the subset of 98 sources that matched the GPSR survey we computed spectral indices and performed the KS test to determine if the sources cluster along the Galactic plane.  Figure~\ref{fig:alphahist} is a histogram of spectral indices with an overlayed Gaussian representing what would be expected from a pure extragalactic source population (De Breuck et al.~2000).  Though most $(\sim 90\%)$ of the sources appear to have spectral indices consistent with extra-galactic sources, there is a tail of the distribution towards both flat and steep spectral indices compared to what is expected from a pure extra-galactic population.    Furthermore, the KS test on sources with a GPSR match had a lower significance level ($2.7\sigma$) than the set of all sources, indicating that though some clustering along the plane may exist, the GPSR matched set is more randomly distributed.  We therefore conclude that among the sources with GPSR matches we are seeing a mostly extra-galactic source population with a few ($\sim 10\%$) Galactic sources.   The tail of the distribution towards steep spectral indices are non-thermal Galactic sources (e.g. pulsars, see below), while the tail towards flat spectral indices points towards thermal sources, e.g.~\htwo\ regions and planetary nebulae.

\begin{figure}[p]
\vspace*{-0.75cm}
\begin{center}
\mbox{\psfig{file=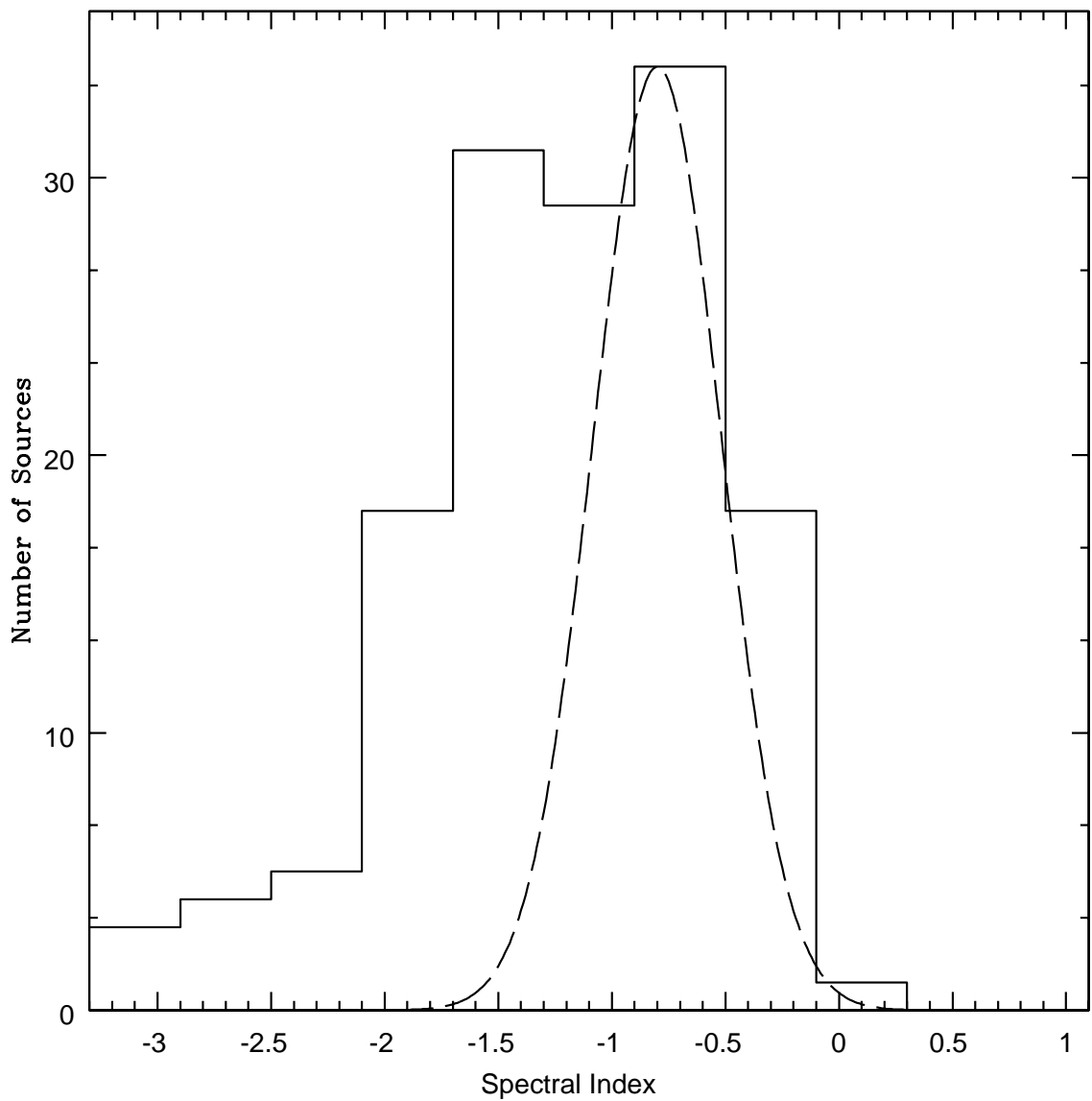,angle=0,width=1.0\textwidth,silent=}}
\end{center}
\vspace*{-1cm}
\caption[]{$\alpha_{1.4}^{0.33}$ spectral index ($S \sim \nu^{\alpha}$) histogram of the 143 sources not detected in the GPSR 1.4 GHz survey (Zoonematkermani et al.~1990 \& Helfand et al.~1992) assuming all sources have a 1.4 GHz flux density of 10 mJy.  The detection threshold of the GPSR survey in the area of interest is 5-10 mJy depending on position.  Note that since the 20 cm flux density is an upper limit, the spectral indices are upper limits, i.e. they may all be steeper than displayed here.  The Gaussian represents the distribution expected if the sources were purely extra-galactic (De Brueck et al.~2000).}
\label{fig:nogpsr_alphahist}
\end{figure}

Because the GPSR has such well matched resolution and sensitivity to our image, the remaining 143 small diameter sources ($\sim$60\%~of our small diameter population)are of interest because they must represent a non-thermal population of sources of relatively steep spectral index.  In the area of interest the GPSR has a detection threshold of 5--10 mJy.  Figure~\ref{fig:nogpsr_alphahist} is a spectral index histogram of sources without a GPSR match assuming a 1.4 GHz flux density of 10 mJy.  Assuming this flux density value results in spectral indices which are upper limits, i.e. all sources must have a spectral index at least this steep.  Again overlayed is a Gaussian representing what is expected from a purely extra-galactic population (De Breuck et al.~2000).  While roughly 50\% of the sources in the region have spectral index upper limits consistent with an extra-galactic population and could be background radio galaxies falling below the sensitivity limit of the GPSR, a large number of the sources are far too steep to be consistent with being background radio galaxies.  Moreover the KS test ruled out a null hypothesis with a $5.8\sigma$ confidence, indicating that these sources tend to strongly cluster along the Galactic plane.  Hence we believe that we are detecting a population of steep spectral index sources of Galactic origin.  Hypothesis for the identity of these sources are pulsars (\S\ref{subsec:pulsars}), stellar clusters or young stellar objects (\S \ref{subsec:yso}), and young Galactic SNRs (\S \ref{subsec:snr}).

\subsubsection{Pulsars and Pulsar Candidates}\label{subsec:pulsars}

Current periodicity searches for pulsars are well known to be biased against short-period, distant, highly dispersed or scattered, and tight binary pulsars (as the recent Lyne et al.~2004 detection of J0737$-$3039 illustrates) (e.g.~Cordes \& Lazio 1997).  As pulsars are expected to have a small diameter and a steep spectral index, this high-resolution, low frequency survey might be expected to be a more effective tool for finding Galactic center pulsars.  For this reason, we have examined our catalog for possible pulsar candidates.

Table~\ref{table:pulsars} lists four previously known pulsars which were detected by checking our small diameter sources against the ATNF pulsar database\footnote{$\langle$http://www.atnf.csiro.au/research/pulsar/psrcat/$\rangle$} (Manchester et al.~2003).  Table \ref{tab:missedpulsars} lists ten known pulsars in the search area that were not detected.  Low flux density at higher frequencies and high image RMS due to positions far from the phase center are consistent with these non-detections with the exception of B1737$-$30.  In the case of this source, higher frequency detections ($\nu > 1$~GHz;~Lorimer et al.~1995 \& Taylor Manchester \& Lyne~1993) indicate that the source could be marginally detected in our image.  However 
this pulsar appears to have a spectral index turnover at frequencies below $\sim 600$~MHz (D. Lorimer 2004, private communication) which would explain this non-detection.  Other previously known sources at comparable distances from the phase center whose higher frequency flux densities suggested they appear at 330 MHz were detected at expected levels.  Hence the non-detection of B1737$-$30 is most likely not a sensitivity related issue.

We identify 30 sources as pulsar candidates based on the following criteria:  candidates must have a small diameter (deconvolved size $< 15\arcsec$ along the major axis) and either have a steep spectrum ($\alpha^{1.4}_{0.33} \le -1.0$) or their non-detection in the Columbia 1.4 GHz survey (Zoonematkermani et al.~1990 \& Helfand et al.~1992) implies a steep spectrum.  The compactness criteria is motivated by the observed diameter of \sgrastar\ at this frequency ($\sim 13\arcsec$; Nord et al.~2004) and is set at a size greater than the beam size due to the scattering discussed in the previous section.  Table \ref{table:pcandidates} gives the details of the pulsar candidates.

Since we have no frequency-time information on any of these sources, we cannot say how many, if any, are pulsars.  However, sources on this list that are not pulsars are still interesting sources and require follow on observations.  

\subsubsection{Steep Spectrum Stellar Clusters and Young Stellar Objects}\label{subsec:yso}

In a previous paper based partly on this image (Yusef-Zadeh et al.~2003), the low frequency emission of the Arches stellar cluster (G0.121$+$0.017, GCPS G0.123$+$0.017 in this survey) was examined.  This stellar cluster is a compact, thermal source at frequencies between 1.4 and 8 GHz, but is strongly non-thermal between 0.33 and 1.4 GHz ($\alpha^{1.4}_{0.33}=-1.2\pm0.4$).  The mechanism for this non-thermal emission is hypothesized to be colliding wind shocks where the stellar winds of mass losing stars collide with the ambient medium.  Of the other two known young stellar clusters in the Galactic center region, the Quintuplet is undetected and the central cluster is undetectable due to its proximity to the Sgr~A supernova remnant.  We note that the SIMBAD matches in Table~\ref{tab:catalog} have 13 sources within 1\arcmin\ of young stellar objects (YSO).  It is possible that part of the population of steep, Galactic sources discussed in \S\ref{subsec:psourceclass} is comprised of such objects.

\subsubsection{Young Galactic Supernova Remnants}\label{subsec:snr}

Though six Galactic supernovae have been detected in the last 1000 years (Clark \& Stephenson 1977), between 20 and 40 are thought to have occurred (one per $40\pm10$~yrs; Tammann, L\"{o}effler, \& Schr\"{o}eder 1994).  The question of the missing SNRs was statistically addressed by Green~(1991) by noting two main detection biases; SNRs must have enough surface brightness to be detected, but also must have an angular extent more than several times the beam size in order to be identified.  This results in a bias toward detection of extended, presumably older SNRs.  As SNRs are non-thermal in nature and cluster strongly along the Galactic plane, a low-frequency survey for small diameter radio sources in the Galactic center such as this one could be ideal for identifying a missing young remnant population. 

Assuming an expansion rate of 2000 km~sec$^{-1}$, a 1000 year old remnant would have a diameter of 2~pc, or $\sim$45\arcsec\ at an assumed galactocentric distance of 8~Kpc.  Indeed, one such compact remnant, G1.18$+$0.33 with a diameter of $\sim 1\arcmin$ is detected in this image and is discussed in \S\ref{snr118p033}.  A remnant only 300 years old or 24 Kpc in distance could easily be classified as a small diameter source ($\lesssim 25\arcsec$) in this survey (Table~\ref{tab:catalog}).

Actual identification of young SNRs from our small diameter source list is difficult.  SNRs typically have a spectral index of $-0.7<\alpha<0.0$, making identification by spectral index difficult in a field dominated by background radio galaxies of nearly the same spectral index range.  SNRs can be significantly polarized, but depolarization by intervening ISM makes polarization work at 330 MHz difficult.  The only possible identifier is morphology.  The small diameter sources were scrutinized by eye for evidence of shell structure but no objects were identified in this manner.  Even this identifier may be biased against identifying plerion-type SNRs.  Thus while some of these sources may be young Galactic SNRs, we have no way of identifying them from among our detected small diameter sources.

\subsection{Evidence for a Scattering Screen}\label{sec:scattering}

Interstellar free electron scattering towards the Galactic center is known to be both large (van Langevelde et al~1992; Frail et al.~1994; Lazio \& Cordes 1998a,b; Bower et al.~1999) and potentially spatially variable (Lazio et al.~1999).  The recent NE2001 model (Cordes \& Lazio 2004) describes the scattering toward the Galactic center by a smoothly distributed screen as well as areas of strong scattering needed to predict large and/or anomalous scattering towards certain sources.  The expected amplitude of angular broadening for a Galactic source seen through this screen is approximately 12\arcsec, based on the diameters of \sgrastar\ and various OH masers when scaled to 330 MHz.  Sources closer than the Galactic center will have smaller scattering diameters while more distant objects will be more heavily broadened, potentially by a large amount.  The angular broadening of an extragalactic source may range from small (less than our beam diameter) to extremely large (many arcminutes), depending upon the porosity of the scattering screen.

\begin{figure}[p]
\vspace*{-0.75cm}
\begin{center}
\mbox{\psfig{file=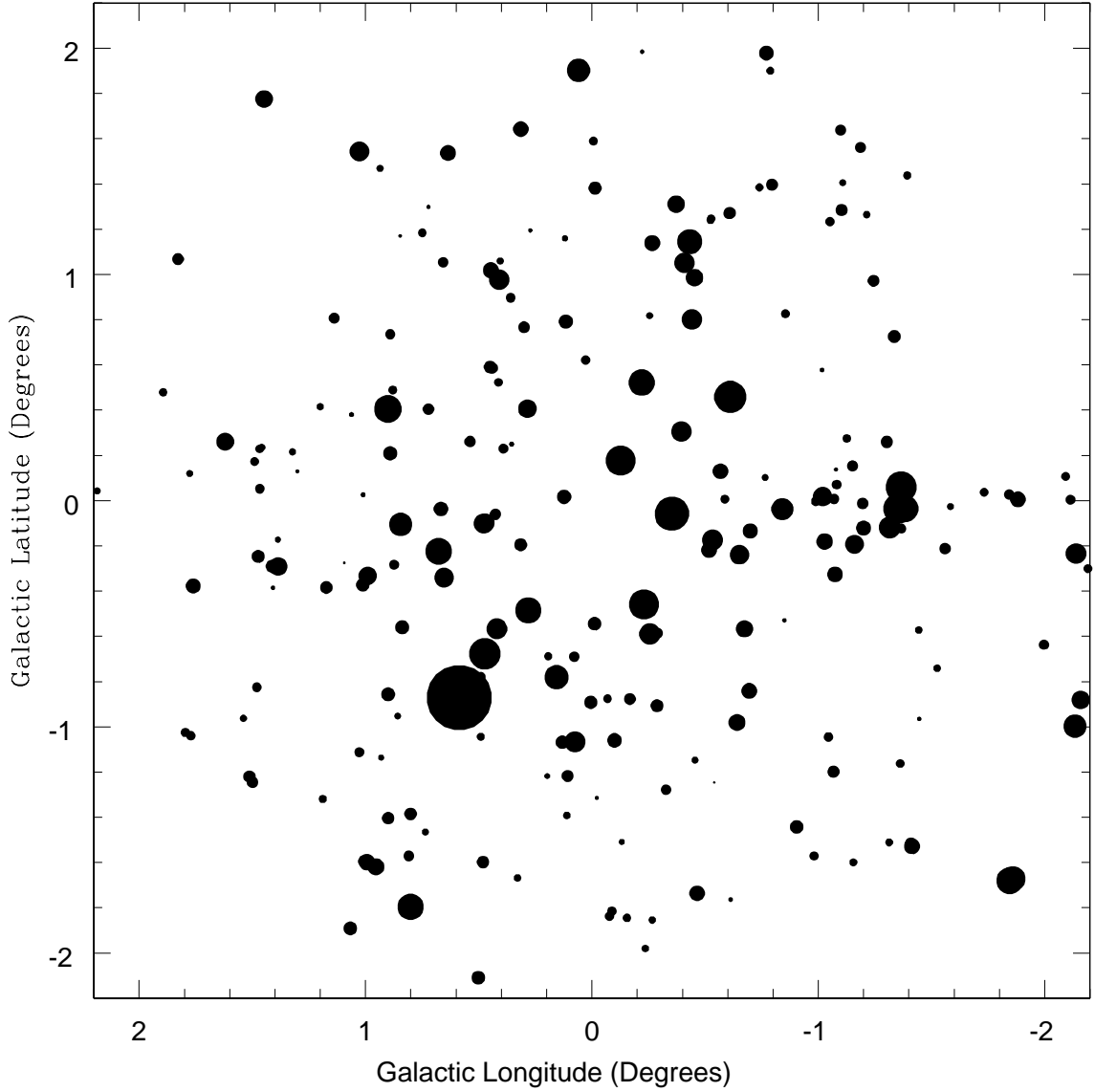,angle=0,width=1.0\textwidth,silent=}}
 \end{center}
\vspace*{-1cm}
\caption[]{Positions and relative deconvolved sizes of all small diameter sources in Galactic coordinates.  No correlation between distance from the Galactic plane and the deconvolved source size was detected.}
\label{fig:lb_sizes}
\end{figure}

Our survey is at a single frequency, so we cannot attribute the diameter of our sources exclusively to interstellar scattering as intrinsic source structure may also contribute.  None the less, the spatial density of sources in our survey combined with its relatively low observation frequency means that scattering effects might still be identified in a statistical sense.
Figure~\ref{fig:lb_sizes} shows the small diameter sources in Galactic coordinates with their relative deconvolved sizes.  We have checked for a correlation between source diameter and source position -- both as a function of distance from the Galactic plane and as a function of distance from the phase center.  We detect no correlation.

However, we do think that we are detecting the signature of the hypothesized scattering screen in differential source counts.  Figure~\ref{fig:dnds} clearly shows source counts fall off strongly towards lower flux density.  Angular broadening conserves flux density, but sources are detected via their maximum intensity, which will decrease as the square of the diameter of the source.  For instance a source with an intrinsic diameter of 10\arcsec\ will have its maximum intensity decreased by $\sim$45\% if broadened by 2\arcsec.  A source that would have been just at the detection limit would become undetectable.  A source with higher intrinsic intensity would still be detectable, and its flux density would remain unchanged.

\subsection{Steep Spectral Index Sources}\label{subsec:steep}

Sensitive low frequency observations are ideal for finding steep-spectra sources.  Several sources in our survey with cross identifications in Table~\ref{tab:crossid} have measured spectral indices that are
very steep ($\alpha^{1.4}_{0.33} \leq -1.8$) and therefore require scrutiny.  These sources are discussed below.

\begin{description}
\item[GCPS G359.535$-$1.736]This source has a spectral index of $\alpha^{1.4}_{0.33} = -1.9$, a deconvolved size of $10.6\arcsec \times 3.0\arcsec$, and a position angle of 102\arcdeg.  A large length to width ratio and a position angle significantly different from that of the clean beam suggests that this source may be an unresolved radio galaxy with only a single component in the 1.4 GHz survey, akin to GCP0.131$-$1.068.

\item[GCPS G0.131$-$1.068]This source is identified with two sources at 1.4 GHz (GPSR G0.131$-$1.065 \& GPSR G0.131$-$10.67), with a spectral index of $\alpha^{1.4}_{0.33} = -1.4$ and $-2.0$ respectively.  The 330 MHz source is quite elongated with a deconvolved size of $11.8\arcsec \times 3.5\arcsec$, and has a position angle of 60\arcdeg, significantly different than the position angle of the beam. If the flux density of the two 1.4 GHz sources are added, the resulting spectral index is $-1.1$.  This source is almost certainly a blending of the two lobes of a radio galaxy.

\item[GCPS G0.993$-$1.599]This source has a spectral index of $\alpha^{1.4}_{0.33} = -2.0$, has a size of 22\arcsec $\times$ 10\arcsec\ with a position angle of 4\arcdeg, and appears slightly diffuse.  If this source is associated with GPSR G1.003$-$1.594 (30\arcsec\ to the north), the two sources would have the morphology of an FR~II radio galaxy.  A potential difficulty with this classification is the steep and quite different spectral indeces ($\alpha^{1.4}_{0.33} = -2.0$ for GPSR G0.993$-$1.599 versus $\alpha^{1.4}_{0.33} = -1.0$ for GPSR G1.003$-$1.594.

\item[GCPS G1.027$+$1.544]This source has a size of 26\arcsec $\times$ 12\arcsec\ with a position angle of 148\arcdeg\ and appears diffuse.  It is identified with two sources at 1.4 GHz (GPSR G1.025$-$1.545 \& GPSR G1.026$-$1.546).  If the flux densities of both 1.4 GHz sources are used in determining the spectral index, the result is $\alpha^{1.4}_{0.33} = -1.4$.  While morphologically this source appears to be an extragalactic source, its spectral index remains steeper than is common for extragalactic sources.

\item[GCPS G1.540$-$0.961]This source is a known pulsar PSR B1749$-$27.  We derive a $\alpha^{1.4}_{0.33}$ spectral index of $-3.0$ for this source.
\end{description}

\subsection{Sagittarius A$^{*}$}\label{sec:sgrastar}

Sagittarius A$^{*}$, the radio source associated with our Galaxy's central massive black hole, was detected utilizing a subset of this data.  This is the first detection of this source at comparable frequencies, and the lowest frequency detection to date.  This detection, as well as implications for emission mechanisms and the location Sagittarius A$^{*}$ with respect to other objects in the Sagittarius~A region are detailed in Nord et al.~(2004).

\section{Extended Sources}\label{sec:extended}

Given the relatively high resolution of this image, a short discussion on why new extended sources were discovered is warranted.  At a resolution of 45\arcsec\ as in LaRosa et al. (2000), large areas of the GC region are dominated by diffuse flux density which is resolved out in this image.  This allows for these regions to be searched for features that are unresolved or only moderately resolved in one dimension, i.e. NTFs.  Secondly several sources appear unresolved to LaRosa et al., but are now resolved at this higher resolution.

Among the most fascinating of the unique structures in the Galactic Center are the NTFs.  These are remarkably coherent magnetic structures that extend tens of parsecs and maintain widths of only a few tenths of parsecs (e.g., Lang et al.~1999).  It has been hypothesized that the NTFs are part of a globally ordered space filling magnetic field (e.g., Morris \& Serabyn~1996) and if so they would be the primary diagnostic of the GC magnetic field.  An alternative idea is that the NTFs are magnetic wakes formed from the amplification of a weak global field through a molecular cloud-galactic center wind interaction (Shore \& LaRosa 1999)

Though as of yet there is no consensus as to the origin of these structures, they are known to be non-thermal in nature, and therefore high resolution studies at low radio frequencies are important to understanding this phenomenon and for increasing the census of known NTFs.

Nine isolated NTFs were known before this work was completed.  Of those nine, we detect eight as we do not have sufficient surface brightness sensitivity to detect G359.85$+$0.39 (LaRosa, Lazio, \& Kassim, 2001).  Table~\ref{table:oldntfs} summarizes the properties of the previously detected NTFs.  With respect to lower resolution measurements, the filaments tend to have lower flux density and are longer.  Insensitivity to low spatial frequencies is responsible for reducing the overall flux density of the NTFs, but also allows for extracting the fainter ends of the NTFs from the extended flux density near the Galactic Center.  For this reason, the flux density measurements of Table~\ref{table:oldntfs} should be taken only as lower limits.

We report the detection of 20 linear structures, two of which have been confirmed as NTFs (LaRosa et al.~2004).  We regard secure identifications of NTFs as those sources with large length to width ratios which have highly polarized ($\gtrsim$ 10\% linear polarization), non-thermal emission.  With these observations, we can determine only morphology and spectrum where higher frequency observations are available.  Without polarization information, we shall classify the remaining 18 objects as NTF candidates.  Table~\ref{table:newntfs} summarizes the properties of these 20 linear structures.

\begin{figure}[p]
\vspace*{-0.75cm}
\begin{center}
\mbox{\psfig{file=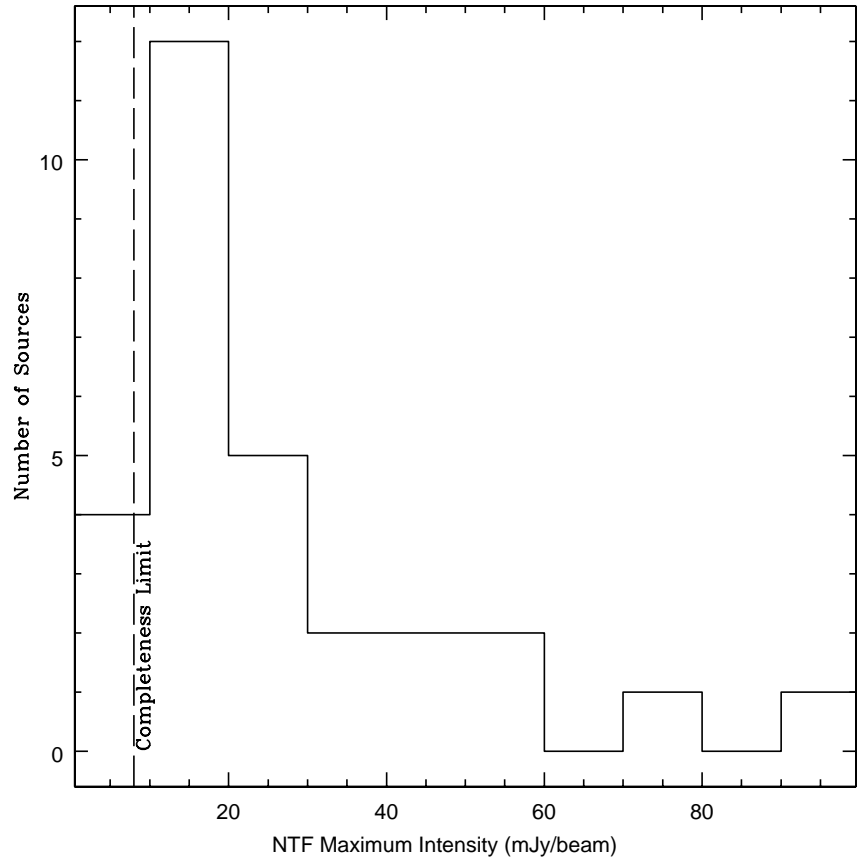,angle=0,width=0.75\textwidth,silent=}}
\end{center}
\vspace*{-1cm}
\caption[]{The histogram of maximum intensity for all known NTFs and NTF candidates.  The vertical dotted line represents our estimate of the minimum intensity at which we can detect NTF candidates reliably.}
\label{fig:ntfhist}
\end{figure}

If we assume that all of these NTF candidates will be eventually confirmed as NTFs, the total number of known NTFs would triple.  Figure~\ref{fig:ntfhist} shows the intensity histogram of NTFs and NTF candidates.  We show the intensity, rather than flux density for two reasons.  First, detections of these sources is based on maximum intensity, not flux density.  secondly baseline subtraction can be difficult for extended sources that pass through regions of diffuse flux density, so the total flux density of the NTFs is uncertain.  Clearly apparent in Figure~\ref{fig:ntfhist} is a rapid increase in the number of potential NTFs at low intensity.  The range of intensities is fairly small, but the increase in number rises faster than linearly with decreasing intensity.  By increasing sensitivity by a factor of $\sim$ 5 over LaRosa et al.~(2000), we have tripled the number of known and suspected NTFs, suggesting that the number of NTFs rises at minimum as $N \sim I^{-0.7}$.  We conclude that just the tip of the NTF luminosity distribution is being detected and we hypothesize that there may be hundreds of low surface brightness NTFs in the GC region.

LaRosa et al.~(2004) discuss the properties of the emerging NTF population in detail, but here we briefly review several noteworthy properties.  Firstly the new NTFs significantly increase the volume of space over which the NTF phenomenon is known to occur.  Though G359.10$-$0.2 remains the furthest southern extent of the NTF phenomenon, new candidate NTFs are now found North, East, and West of those previously known.  The entire population of suspected NTFs now covers $\sim$ 2 square degrees covering Galactic longitudes from $+0.4\arcdeg$ to $-0.9\arcdeg$ and Galactic latitudes from $+0.7\arcdeg$ to $-0.5\arcdeg$.  This observation is of particular importance to NTF models assuming a space filling poloidal field, as the volume over which this field must exist, and therefore the magnetic energy in the field, are now significantly increased.  Of further note is the space distribution of the new candidate NTFs.  Previously G359.10$-$0.2 (The Snake) was the only NTF found south of the Galactic plane, though the Galactic Center radio arc and Sgr C cross the plane.  Thirteen of the twenty  candidate NTFs are south of the Galactic plane.

The orientation of the NTFs is of particular interest due to their potential to discriminate between different NTF origin theories and for tracing Galactic Center magnetic fields.  For the purposes of this discussion, orientation will be defined as the separation angle between the long axis of the NTF and the normal to the Galactic plane.  The nine known isolated NTFs are, with the exception of G358.85$+$0.47, nearly normal to the Galactic plane.  This observation supports the hypothesis that the magnetic field in the region is poloidal in nature (e.g.~Morris \& Serabyn 1996, and references therein).  However, the new NTF population differs significantly with a mean orientation of $35\arcdeg \pm 40\arcdeg$.  Furthermore, NTFs much closer to the plane and to the Galactic Center than G358.85$+$0.47~such as NTF G359.22$-$0.16 are nearly parallel to the Galactic plane.  This suggests that the Galactic Center magnetic field is significantly more complicated than a simple dipole field.  Though it noteworthy that the brightest NTFs align normal to the plane, the new NTF population would appear to imply an larger scale non-poloidal field and/or a disordered component of the magnetic field.  Moreover, the pseudo-random orientation of weaker candidate NTFs may indicate a physical manifestation not directly connected to the properties of any global field.

\subsection{NTFs and NTF Candidates}

\begin{figure}[p]
\vspace*{-0.75cm}
\begin{center}
\mbox{\psfig{file=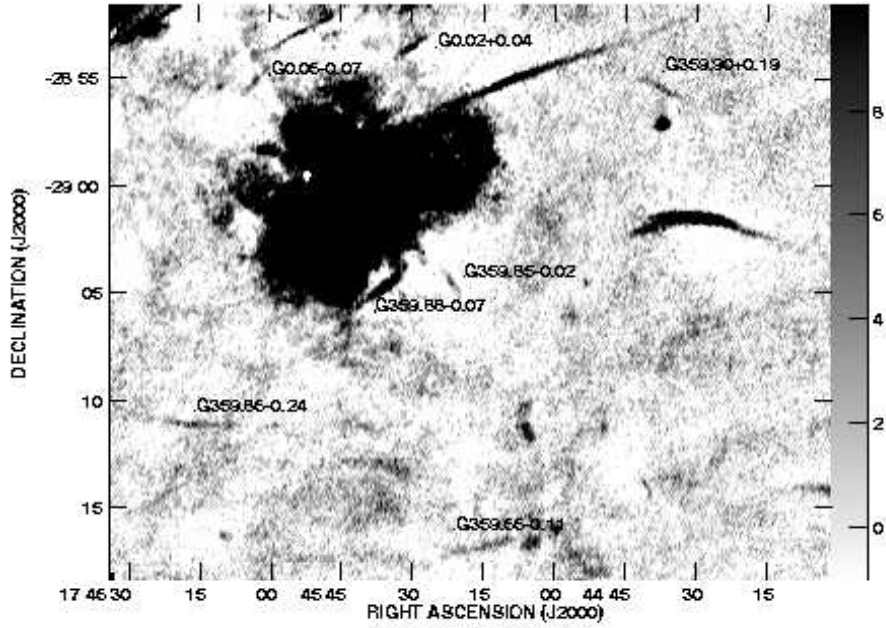,angle=-90,width=0.75\textwidth,silent=}}
\end{center}
\vspace*{-1cm}
\caption[]{Non-Thermal Filaments candidates in the Sgr~A region.  The gray scale is linear between~$-1$ and~10 \mjybm.  Primary beam correction has not been applied but is negligable in this field.}
\label{fig:sgra}
\end{figure}

\begin{figure}[p]
\vspace*{-0.75cm}
\begin{center}
\mbox{\psfig{file=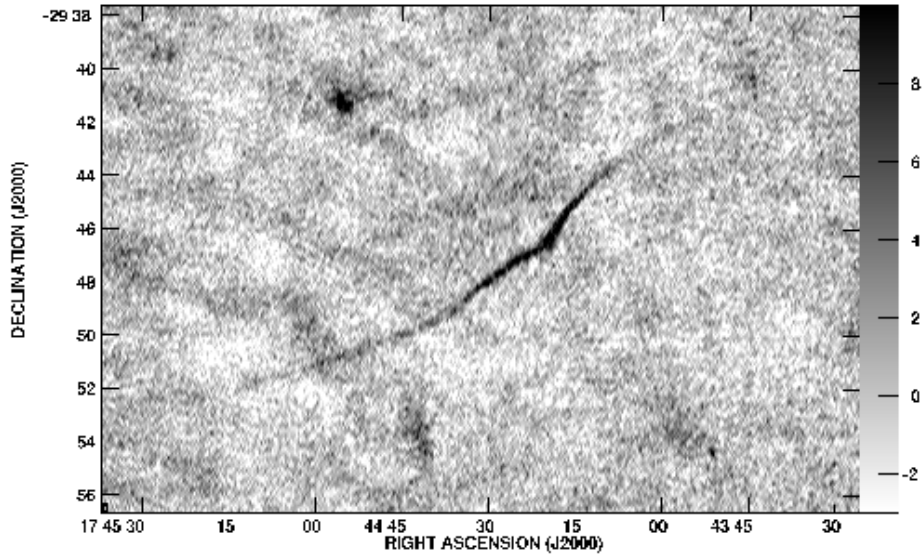,angle=-90,width=0.75\textwidth,silent=}}
\end{center}
\vspace*{-1cm}
\caption[]{Non-Thermal Filament G359.10$-$0.2 (The Snake).  The gray scale is linear between~$-1$ and~10~\mjybm.  Primary beam correction has not been applied; the value at the center of the field is 1.29.}
\label{fig:thesnake}
\end{figure}

\begin{figure}[p]
\vspace*{-0.75cm}
\begin{center}
\mbox{\psfig{file=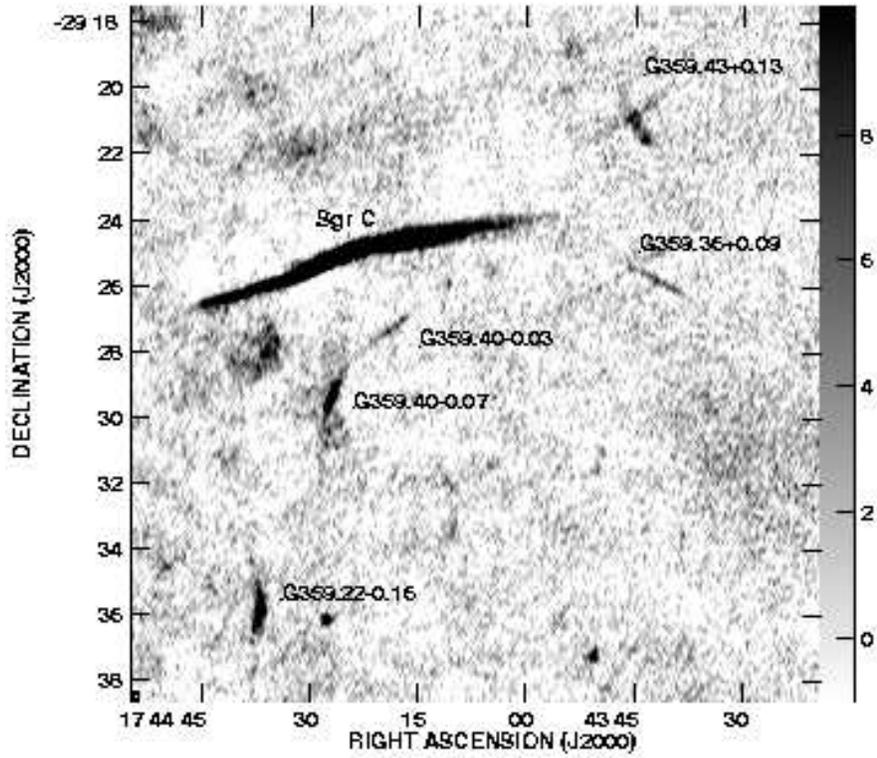,angle=-90,width=0.75\textwidth,silent=}}
\end{center}
\vspace*{-1cm}
\caption[]{Non-Thermal Filaments and candidates in the Sgr~C region.  The gray scale is linear between~$-1$
and~10 \mjybm.  Primary beam correction has not been applied; the value at the center of the field is 1.13.}
\label{fig:sgrc}
\end{figure}

\begin{figure}[p]
\begin{center}
\mbox{\psfig{file=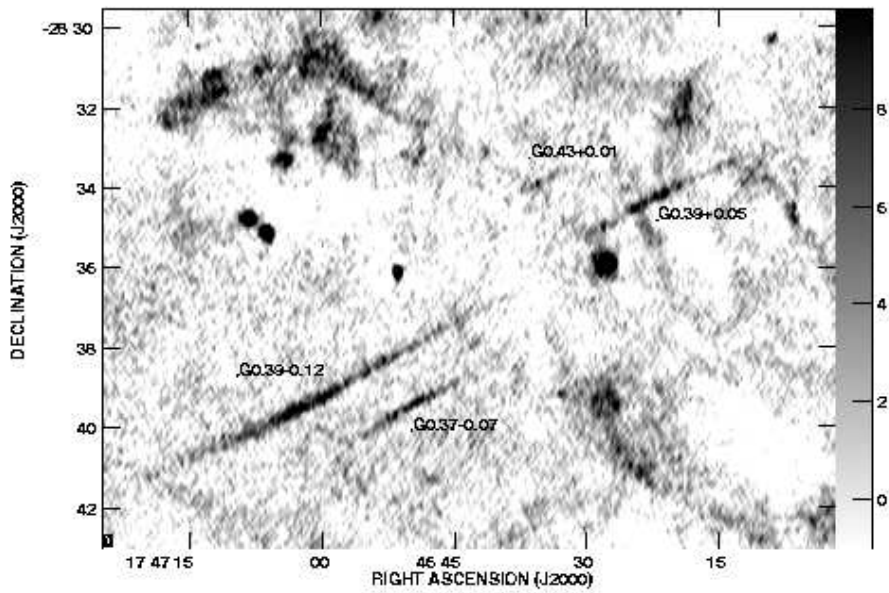,angle=-90,width=0.75\textwidth,silent=}}
\end{center}
\vspace*{-1.5cm}
\caption[]{Non-Thermal Filaments and candidates in the region between Sgr~B1 and the Radio Arc showing three candidate NTFs and one confirmed NTF similar in appearance to the filaments comprising the Galactic Center Radio Arc, which is located approximately 10\arcmin\ to the
south.  The gray scale is linear between~$-1$
and~10~\mjybm.  Primary beam correction has not been applied; the value at the center of the field is 1.08.}
\label{fig:sgrb}
\end{figure}

\begin{figure}[p]
\vspace*{-0.75cm}
\begin{center}
\mbox{\psfig{file=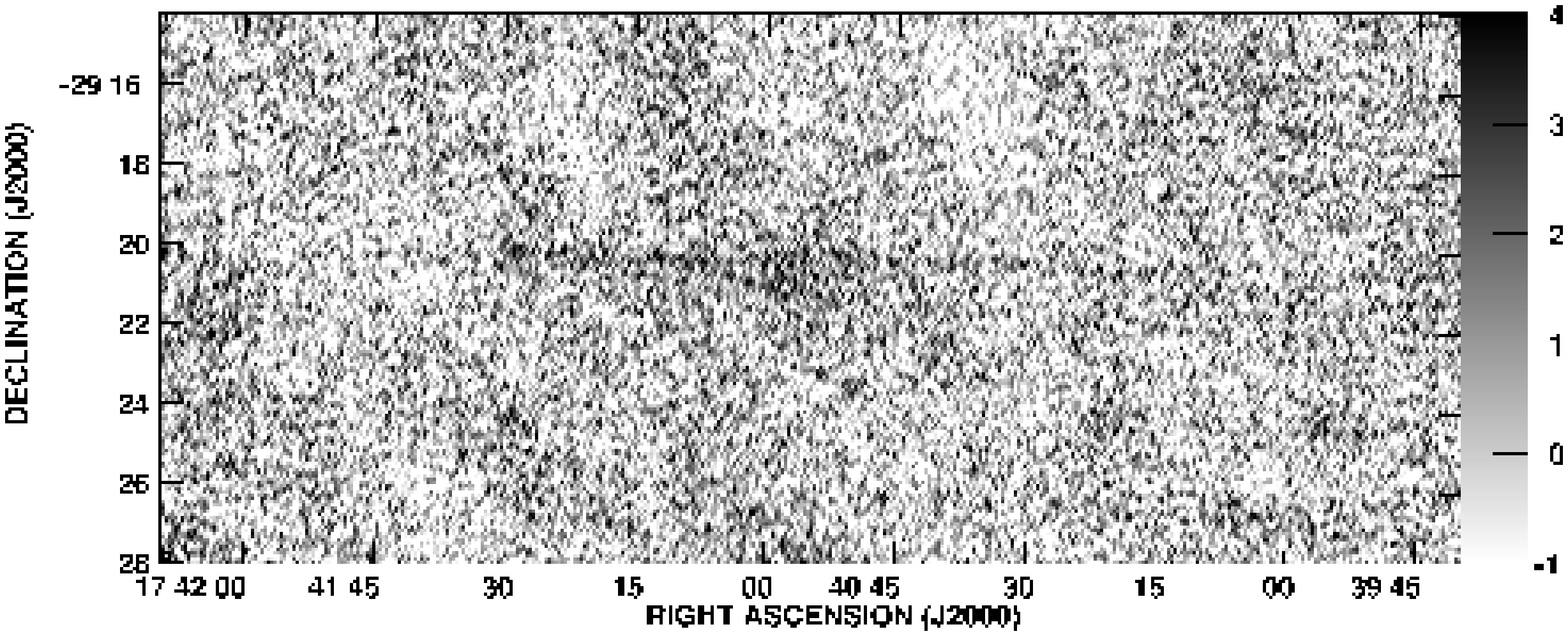,angle=-90,width=0.75\textwidth,silent=}}
\end{center}
\vspace*{-1cm}
\caption[]{Non-Thermal Filament candidate G359.12$+$0.66.  The gray scale is linear between~$-1$ and~4~\mjybm.  This source has an extremely low surface brightness and is best viewed from a distance.  Primary beam correction has not been applied; the value at the center of the field is 1.61.}
\label{fig:359.12+0.66}
\end{figure}

\begin{figure}[p]
\vspace*{-0.75cm}
\begin{center}
\mbox{\psfig{file=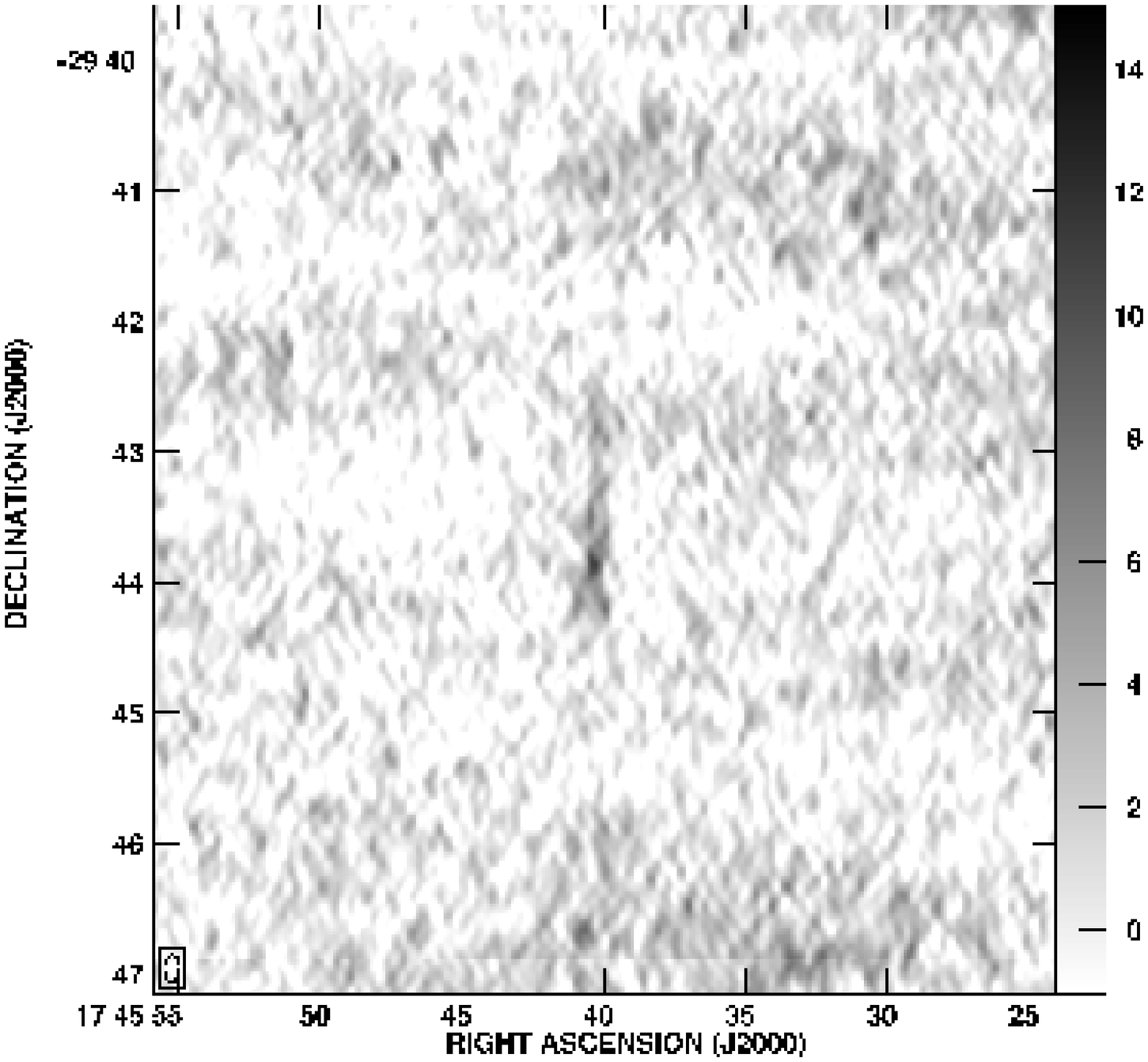,angle=-90,width=0.75\textwidth,silent=}}
\end{center}
\vspace*{-1cm}
\caption[]{Non-Thermal Filament candidate G359.33$-$0.42.  The gray scale is linear between~$-1$ and~15~\mjybm.  Primary beam correction has not been applied; the value at the center of the field is 1.21.}
\label{fig:359.33-0.42}
\end{figure}

\begin{figure}[p]
\vspace*{-0.75cm}
\begin{center}
\mbox{\psfig{file=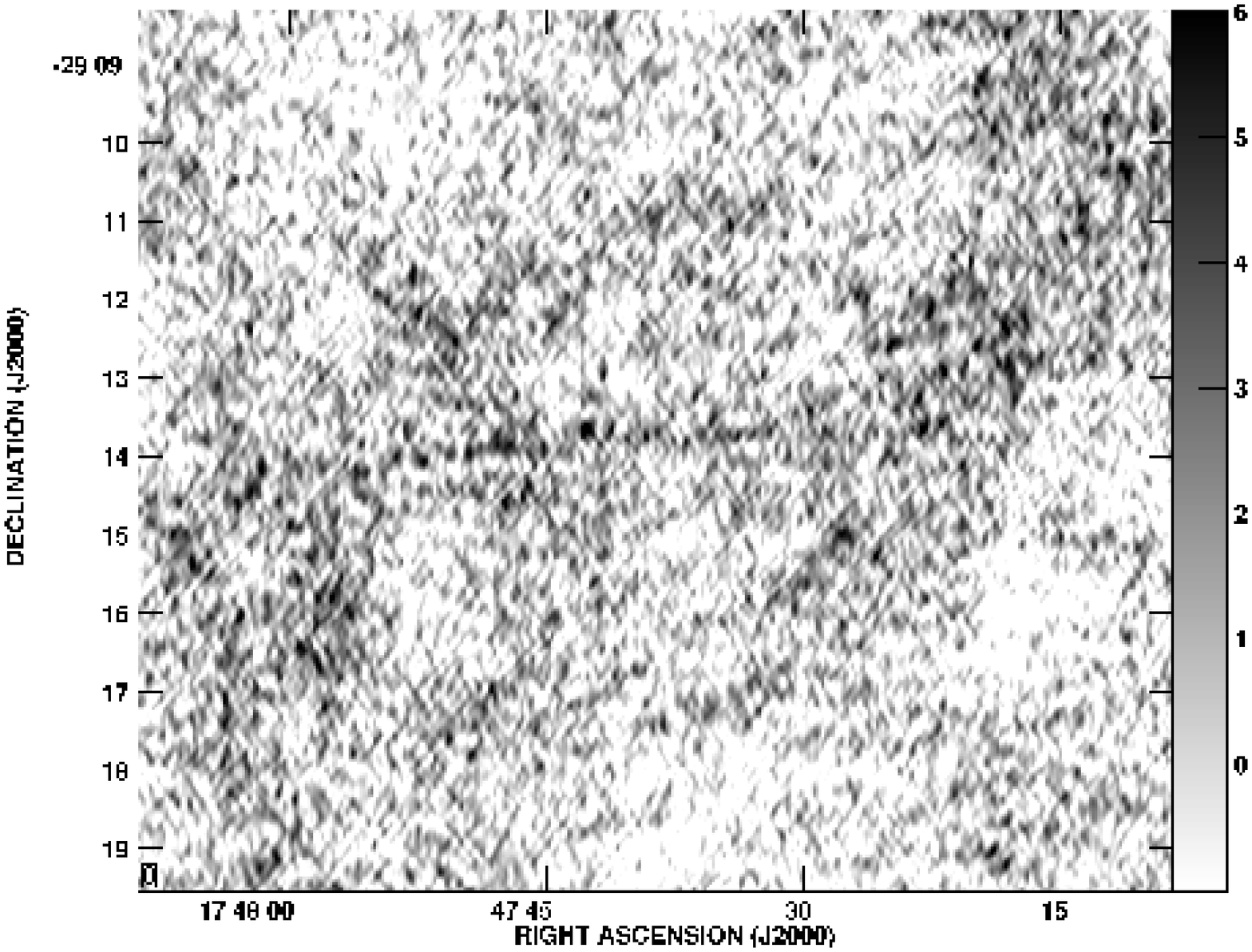,angle=-90,width=0.75\textwidth,silent=}}
\end{center}
\vspace*{-1cm}
\caption[]{Non-Thermal Filament candidate G359.99$-$0.54.  The gray scale is linear between~$-1$ and~6~\mjybm.  This source has an extremely low surface brightness and is best viewed from a distance.  Primary beam correction has not been applied; the value at the center of the field is 1.09.}
\label{fig:359.99-0.54}
\end{figure}

\begin{description}

\item[Candidate NTFs G359.86$-$0.24 and G359.66$-$0.11]Lying to the south of Sgr A in Figure~\ref{fig:sgra}, these long, low brightness NTF candidates both curve northward.

\item[Candidate NTFs G359.88$-$0.07 and G359.85$-$0.02]To the south and west of the supernova remnant Sgr~A East in Figure~\ref{fig:sgra}, and separated by only 4\arcmin, these NTF candidates are nearly perpendicular to each other.  Candidate NTF G359.85$-$0.02 is in itself interesting in that it is so close to the plane and to the Galactic Center, yet is parallel to the plane.  The simple dipole model of the Galactic Center magnetic field would be challenged to explain this NTF orientation.  G359.85$-$0.02 was marginally detected at 620 MHz and labeled "Thread U" in Roy~(2003) and G359.88$-$0.07 was identified in Lang et al.~(1999) as a "streak".

\item[Candidate NTFs G0.02$+$ 0.04 and G0.06$-$0.07]To the north of Sgr A in Figure~\ref{fig:sgra}, these two faint NTF candidates are nearly parallel to the nearby bright filament G0.08$+$0.15.  G0.02$+$0.04 was identified in Lang et al.~(1999) as a "streak".

\item[Candidate NTF G359.90$+$0.19]To the south of the western extension of G0.08$+$0.15, this NTF candidate differs in orientation by roughly 35\arcdeg\ from the nearby bright NTF.  If this candidate is an NTF, it traces what must be significant magnetic field gradient in this region.

\item[NTF G359.10$-$0.2 (The Snake)]G359.10$-$0.2 (Figure~\ref{fig:thesnake}) was reported by LaRosa et al (2000) to have a length of 5.2\arcmin.  Reduced sensitivity to large scale features and increased sensitivity to small scale features in this image show that this feature in fact extends over more than 20\arcmin, has a large 'kink' in the middle, and shows curvature in different directions on each side of the kink; observations which are in agreement with higher frequency observations of this source (Grey et al.~1995).

\item[Candidate NTFs G359.40$-$0.07 and G359.40$-$0.03]To the south of Sgr C in Figure~\ref{fig:sgrc} lies candidate NTF G359.40$-$0.07.  This source was observed by Liszt \& Spiker~(1995) at~18~cm and detected as a small diameter source in LaRosa et al.~(2000).  We derive a 20/90~cm spectral index of $\alpha \approx -0.1$.  
Higher-resolution, 20~cm observations (Lazio \& Cordes~2004) show the source to have a filamentary appearance.

Candidate NTF G359.40$-$0.03 may be a faint extension of G359.40$-$0.07.  If one source, the bright part of is distinctly non-perpendicular to the Galactic plane while the extension curves and becomes more perpendicular.  This demonstrates a significant magnetic field gradient, particularly given its proximity to Sgr~C, which does not.

\item[Candidate NTF G359.36$+$0.09]Figure~\ref{fig:sgrc} shows that G359.36+0.09 lies just to the west of Sgr C.  At high resolution the western end of Sgr~C filament is resolved into
two distinct filaments (Liszt \& Spiker~1995).  The end of the Sgr~C filament begins to flare, and there is linear source G359.36$+$0.09 that may connect to the
bottom filament of \hbox{Sgr~C}.  
A very faint structure appears to  cross this filament between it and Sgr C.  If real, this would be the second example of interacting filaments in the Sgr C region as G359.43$+$0.13 to the north also exhibits a crossing filament.

\item[NTF G359.22$-$0.16]Figure~\ref{fig:sgrc} shows this NTF lies to the south of the eastern part of Sgr~C.  This NTF has been observed to have a polarization of $\sim 40\%$ at 6 cm (LaRosa et al. 2004), confirming this source as an NTF.  This NTF is nearly parallel to the Galactic plane making it only the second confirmed NTF to be parallel to the plane, yet it is much closer to the plane than G358.85$+$0.47 'The Pelican', the only other previously known parallel NTF.
Furthermore, since the end of this source is less than 10 pc in projection from the Sgr C filament, yet nearly normal to Sgr C, a simple dipole Galactic magnetic field structure cannot explain this filament.

\item[Candidate NTF G359.43$+$0.13]Figure~\ref{fig:sgrc} shows this candidate to lie northwest of Sgr C.  This cross-shaped source may be an example of interacting NTFs.
Even if this structure is simply a projection effect we have another
 potential NTF that is parallel to the Galactic plane, and one that is
far closer to Sgr~A than is G358.85$+$0.47, 'The Pelican' (Lang et al.~1999).

\item[NTF G0.39$-$0.12 \& Candidate NTFs G0.37$-$0.07, G0.43$+$0.01, and G0.39$+$0.05]Figure~\ref{fig:sgrb} shows our candidates between Sgr~B1 and the Radio Arc.  They appear to be isolated NTFs that cross the plane with the same orientation as the bundled filaments in the Galactic Center Radio Arc.  NTF0.39$-$0.12 has been observed to have a 6 cm polarization greater than 10\% (LaRosa et al.~2004) and is therefore confirmed as an NTF.  These NTFs are the only known NTFs known north of the Radio Arc and therefore significantly increase the volume over which the NTF phenomena occurs.

\item[Candidate NTF G359.12$+$0.66]Figure~\ref{fig:359.12+0.66} shows the very faint G359.12$+$0.66.  This filament was first detected at higher frequencies (M. Morris 2002, private communication) and is at the limits of detection here.  This is a very long filament ($\sim 15.6\arcmin$) that is far above the Galactic plane, and appears to bifurcate in the middle.

\item[Candidate NTF G359.33$-$0.42]
Figure~\ref{fig:359.33-0.42} shows the short G359.33$-$0.42.  This candidate NTF is very nearly perpendicular to the Galactic plane, making it the only perpendicular candidate south of the Galactic plane.

\item[Candidate NTF G359.99$-$0.54]
Figure~\ref{fig:359.99-0.54} shows the very faint candidate NTF G359.99$-$0.54.  Though no part of this filament has a flux density greater than $3$ times the RMS noise of the image, it is detectable by comparing the flux density of the region to nearby regions.  This filament is the furthest south of the Galactic plane of all the NTFs.

\item[Candidate NTF G359.59$-$0.34]
Figure~\ref{fig:NTF359.59-0.34} shows G359.59$-$0.34.  This short candidate flares significantly to the northwest, more than tripling its width.  Other NTFs are observed to flare, but this is the most extreme example.

\end{description}

\subsection{SNR G1.88$+$0.33}\label{snr118p033}

\begin{figure}[p]
\vspace*{-0.75cm}
\begin{center}
\mbox{\psfig{file=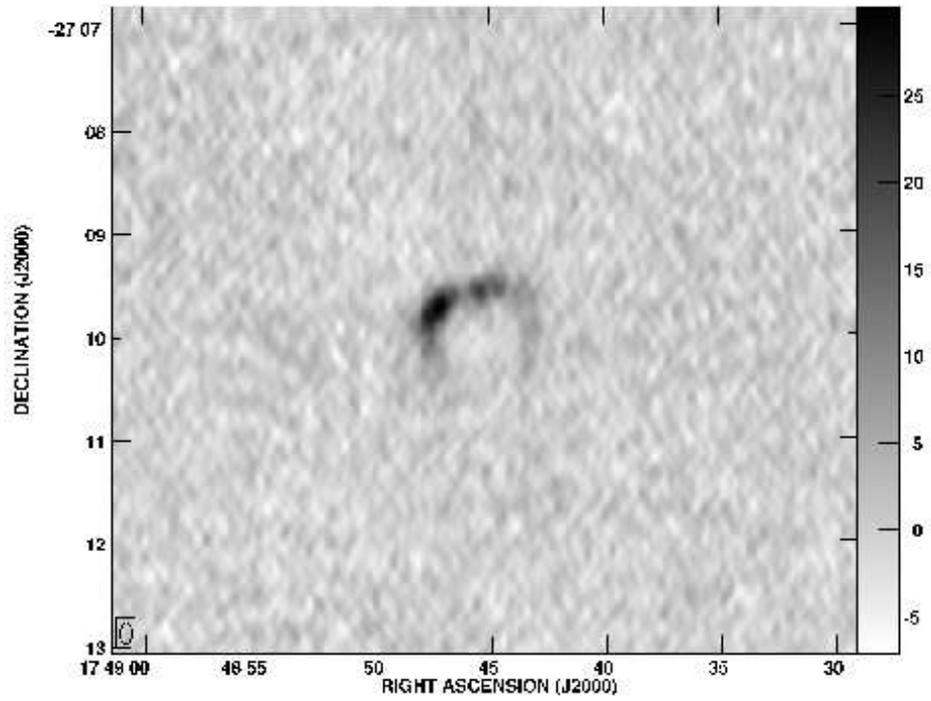,angle=-90,width=0.75\textwidth,silent=}}
\end{center}
\vspace*{-1cm}
\caption[]{Supernova Remnant G1.88+0.33.  The gray scale is linear between~$-7$ and~30~\mjybm.  Primary beam correction has not been applied; the value at the center of the field is 5.01.}
\label{fig:SNR1.88+0.33}
\end{figure}

Figure~\ref{fig:SNR1.88+0.33} shows supernova remnant G1.88$+$0.33. Considered a small diameter source in LaRosa et al. (2000), it is now resolved in our image.  First reported as a SNR in \cite{greenandgull}, this remnant is quite small ($<1\arcmin$), and if it could be shown to be nearby, would be one of the youngest known Galactic supernova remnants.  A 327/74 MHz spectral index of $\approx -0.65$ (Brogan et al. 2004) indicates that the remnant does not have significant 74 MHz absorption.  The fact that the line of sight passes within 2\arcdeg\ of the Galactic center suggests that it may be on the near side of the Galactic center to avoid absorption, but this is not a robust distance indicator.  If indeed it is on this side of the Galactic Center ($< 7.8$ Kpc) it would be less than 3 pc in diameter, smaller than 428 year old Tycho \citep{1995A&A...299..193S}.

\subsection{G0.4$-$0.6} \label{sec:hyman}

\begin{figure}[p]
\vspace*{-0.75cm}
\begin{center}
\mbox{\psfig{file=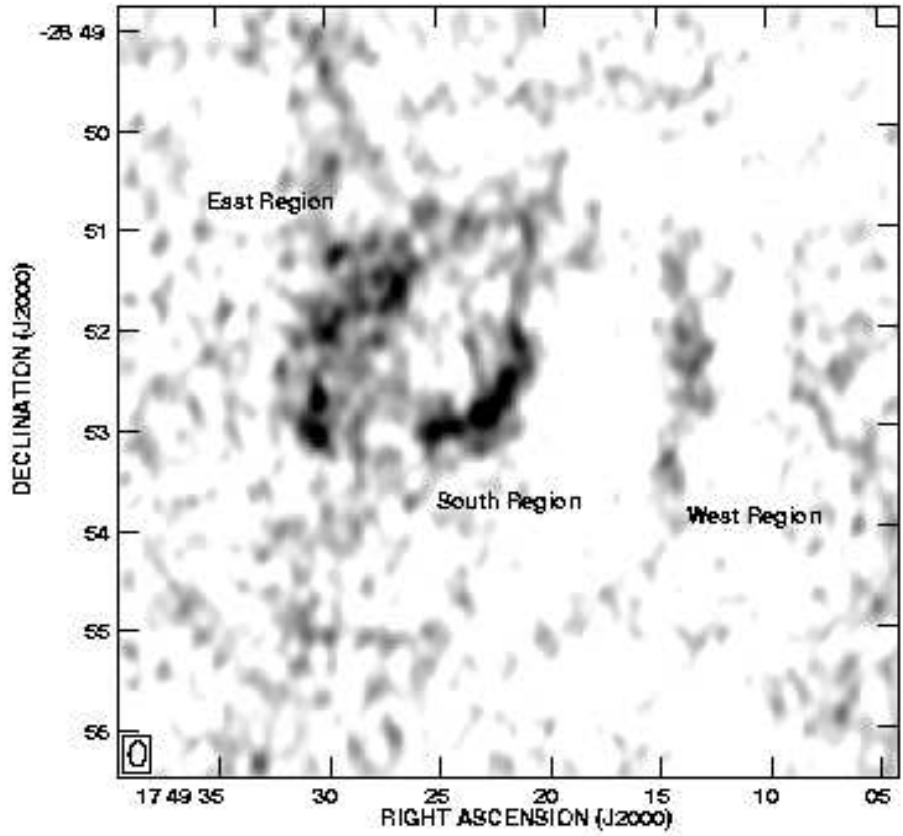,angle=-90,width=0.75\textwidth,silent=}}
\end{center}
\vspace*{-1cm}
\caption[]{\htwo\ region G0.4$-$0.6.  The gray scale is linear between~$0$ and~10~\mjybm.  Primary beam correction has not been applied; the value at the center of the field is 1.28.  The three regions indicated were used to determine spectral indices as described in Section \ref{sec:hyman}.}
\label{fig:hymanHII.ps}
\end{figure}

G0.4$-$0.6 is a moderately strong ($1-2$ Jy), extended ($\sim$ 5\arcmin) region of
emission, shown in Figure \ref{fig:hymanHII.ps}, and located approximately one degree east
of the Galactic center. Its morphology on the much lower resolution LaRosa et al.~(2000) 330 MHz image consists of an
incomplete spherical shell with a central component of emission, resembling a composite SNR. However, a comparison with
an earlier VLA image at 1.6 GHz, kindly provided by H. Liszt, indicates
that each of the source components shown in Figure \ref{fig:hymanHII.ps}\ has a flat to inverted spectrum. VLA
observations obtained by us in 2001 at 4.8 GHz, together with previous,
lower resolution, single dish measurements at 4.9 and 10 GHz (Altenhoff
et al. 1979; Handa et al. 1987), demonstrate that the spectrum is flat
at high frequencies.

An upper limit of only 1\% polarization was determined from the recent
4.8 GHz observation. We consider unlikely the possibility of severe
depolarization due to a foreground thermal plasma. The morphology at 4.8
GHz suggests that the region of emission may actually consist of two
physically distinct sources with one comprised of the eastern and
southern regions as indicated in the figure, and a second, separate
western region.

Based on its spectrum and unpolarized emission, we conclude that
G0.4-0.6 is most likely an \htwo\ region(s) with decreasing flux density below $\sim$ 1
GHz due to self-absorption. This interpretation is supported by a single
dish observation (Downes et al. 1980) that detected H110a and H$_2$CO
recombination lines from the region.  Also, their radial velocity
measurements provide evidence that the source is located in the GC.




\section{Conclusions}\label{sec:conclude}

We have a presented a high resolution, high sensitivity image of the Galactic Center at 330 MHz.  Synthesized from new observations and utilizing improved low frequency data reduction procedures, this image improves on previous GC 330 MHz images (LaRosa et al.~2000) by roughly a factor of five in both resolution and surface brightness sensitivity.  

In this image we have identified 241 small diameter sources (diameters $\lesssim$ 15\arcsec), tripling the number detected in previous low-frequency images of this region (LaRosa et al.~2000).  Of these, roughly 40\% can be identified with sources detected at higher frequencies, primarily those in the 1.4 GHz GPSR catalog (Zoonematkermani et al.~1990 \& Helfand et al.~1992), enabling spectral index determinations.  The spectral index distribution is broadly consistent with that expected from an extragalactic population of sources, though there are significant tails to both steep spectrum and flat spectrum sources.  The remaining $\sim$60\% show clustering along the Galactic plane and roughly 50\% of this population have spectral index upper limits ($\alpha \leq -0.7$) which are inconsistent with extra-galactic sources.  The exact nature of these sources is unknown, but candidates include young SNRs, pulsars, and stellar wind shocks from young stellar objects and/or stellar clusters.  A paucity of low flux density small diameter sources with respect to an extra-galactic population is interpreted as the effect of a free electron scattering screen along the Galactic plane.

Of fourteen known pulsars in the survey area, four are detected.  Non-detections are explained through low intrinsic brightness at higher frequencies and/or positions far from the phase center of the observations.  Thirty sources were classified as pulsar candidates based on morphology and spectrum.

We have identified twenty non-thermal filaments and NTF candidates.  If all are eventually confirmed, the census of NTFs in the Galactic center will have tripled.  The pseudo-random orientation of these filaments is in stark contrast to previously detected filaments, which with one exception are all nearly normal to the Galactic plane.  As NTFs have been thought to be tracers of the Galactic center magnetic structure, the introduction of randomly oriented filaments necessitates re-examination of the paradigm of a strong, ordered, global magnetic field, currently accepted theories of NTF formation, or both.

Future work will include a 330~MHz Galactic Center image utilizing all VLA configurations in combination with data from the Green Bank Telescope, and 74 MHz VLA imaging of the Galactic center.



\acknowledgements{

The original data request was written with the help of K. Anantharamaiah.  "Anantha" passed away during the initial stages of this project and will be missed greatly.

The authors would like to thank Mariana S. Lazarova and Jennifer L. Neureuther, students at Sweet Briar College for their assistance in small diameter source location and quantification.

Basic research in radio astronomy at the NRL is supported by the Office of Naval Research.  S.~D.~H.~was supported by the Jeffres Memorial Trust and Research Corporation.   The National Radio Astronomy Observatory is a facility of the National Science Foundation operated under cooperative agreement by Associated Universities, Inc.  This research has made use of the SIMBAD database, operated at CDS, Strasbourg, France}




\clearpage

\clearpage

\clearpage

\clearpage

\clearpage

\begin{figure}[p]
\begin{center}
\mbox{\psfig{file=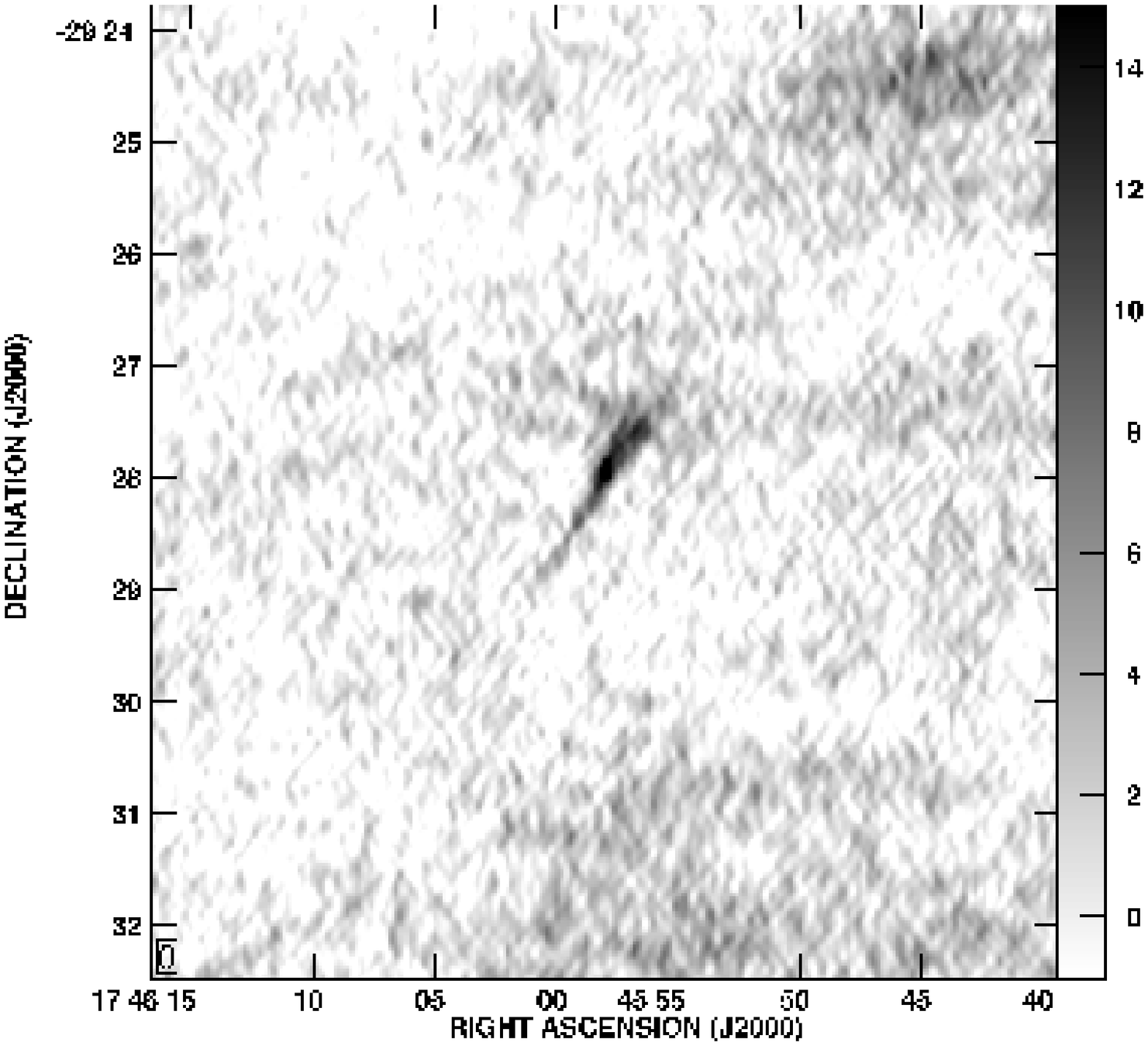,angle=-90,width=0.75\textwidth,silent=}}
\end{center}
\vspace*{-1.5cm}
\caption[]{Non-Thermal Filament candidate NTF~359.59$-$0.34.  The gray scale is linear between~$-1$
and~15~\mjybm.  Primary beam correction has not been applied; the value at the center of the field is 1.08.}
\label{fig:NTF359.59-0.34}
\end{figure}

\clearpage

\clearpage


\begin{deluxetable}{lccccc}

\tabletypesize{\small}

\tablewidth{0pc}

\tablecaption{Observational Summary}

\tablehead{\colhead{Date} & \colhead{Configuration} & \colhead{$\nu$ (MHz)} & \colhead{$\Delta \nu$ (MHz)}& \colhead{Integration(hours)} & \colhead{Beam}}

\startdata

24 October 1996&A&332.5&6&5.55&$9\arcsec \times 5\arcsec$\\

4 March 1998&A&327.5&3&5.38&$9\arcsec \times 5\arcsec$\\

26 September 1998&B&327.5&3&5.47&$36\arcsec \times 20\arcsec$\\

\enddata

\label{tab:observe}

\end{deluxetable}

\begin{deluxetable}{llllcccccl}

\tabletypesize{\footnotesize}
\tablecaption{Galactic Center P-band Survey (GCPS) of Small Diameter Sources \label{tab:catalog}}

\tablecomments{I is intensity, RMS is the local sensitivity, S is flux density, $\theta$ is the average deconvolved size,  Offset is the distance of the source from the phase center (J2000 $17^{\mathrm{h}}45^{\mathrm{m}}40.045^{\mathrm{s}} -29\arcdeg 00\arcmin 27.900\arcsec$).}
\tablerefs{Matches made using the SIMBAD database, operated at CDS, Strasbourg, France $\langle$http://simbad.u-strasbg.fr/Simbad$\rangle$}

\tablehead{

\colhead{\#}       &\colhead{Name}              & \colhead{RA}               & \colhead{Dec}         &
\colhead{$I$}      & \colhead{RMS}              & \colhead{$S$}         & 
\colhead{$\theta$} & \colhead{Offset}           & \colhead{SIMBAD\tablenotemark{a}}\\ 

&& \multicolumn{2}{c}{(J2000)} 
& \multicolumn{2}{c}{(\mjybm)}  
& \colhead{(mJy)}                       & \colhead{(\arcsec)}         
& \colhead{($\arcmin$)}                 & \colhead{Match}
}

\startdata

1  &357.796-0.790 & 17 43 24.22 &-31 13 39.6 & 89.7 & 13.2 & 234.2 & 10.4 & 136.2 & 	\\
2  &357.809-0.300 & 17 41 29.34 &-30 57 32.6 & 168.8 & 11.0 & 237.2 & 5.8 & 129.0 & 	\\
3  &357.841-0.881 & 17 43 52.51 &-31 14 13.0 & 82.6 & 13.1 & 337.0 & 12.0 & 136.0 & 	\\
4  &357.861-0.234 & 17 41 21.34 &-30 52 44.8 & 64.4 & 9.9 & 91.6 & 13.5 & 125.3 & 	\\
5  &357.866-0.997 & 17 44 23.82 &-31 16 36.2 & 233.4 & 13.5 & 881.6 & 15.2 & 137.0 &	\\
6  &357.886+0.004 & 17 40 28.64 &-30 43 56.7 & 186.1 & 9.3 & 292.4 & 6.6 & 123.2 &	\\
7  &357.907+0.107 & 17 40 7.36  &-30 39 36.4 & 84.6 & 9.2 & 129.0 & 5.7 & 122.6 & 		\\
8  &358.003-0.637 & 17 43 18.22 &-30 58 17.1 & 427.5 & 9.0 & 654.5 & 6.6 & 121.9 & 	\\
\enddata
\tablenotetext{a}{Within 1\arcmin:  PSR, puslar; IRAS, IRAS infrared source; YSO, young stellar object; X, X-ray object; PN, planetary nebula;PN?, possible planetary nebula; IR isogal infrared source; XRB, x-ray binary; \htwo, \htwo\ region; OPC, open cluster; MC, molucular cloud; SY1, Seyfert 1 galaxy.  Stellar matches were neglected, and radio source matches are in Table~\ref{tab:crossid}.}
\tablenotetext{b} {Transient source, see Hyman et al.~2002.}

\end{deluxetable}

\begin{deluxetable}{lccrrrrrr}

\tabletypesize{\scriptsize}
\tablecaption{Radio Cross-identifications \label{tab:crossid}}
\tablerefs{(GPSR) Zoonematkermani et al.~1990 \& Helfand et al.~1992,~(2LC) Lazio \& Cordes 1998,~(GPSR5) Becker et al.~1994}
\tablehead{

 \multicolumn{3}{c}{} \\

  & \colhead{S}   & \colhead{Angular} &

 \colhead{GPSR} & & 

 \colhead{2LC} &  & 

 \colhead{GPSR5} & \\ 

 \colhead{Name} & \colhead{(mJy)}   & \colhead{Diameter(\arcsec)} &

 \colhead{Name} & \colhead{$\alpha^{1.4}_{0.33}$} & 

 \colhead{Name} & \colhead{$\alpha^{1.4}_{0.33}$} & 

 \colhead{Name} & \colhead{$\alpha^{0.33}_5$}} 

\startdata

357.796$-$0.790 &   234.2 &  10.40 & 357.795$-$0.788 &  $-$1.3 &                 &         &                   &         \\
357.809$-$0.300 &   237.2 &   5.79 & 357.810$-$0.298 &  $-$0.7 &                 &         & 357.808$-$0.299   &  $-$0.9 \\
357.841$-$0.881 &   337.0 &  12.03 & 357.840$-$0.880 &  $-$1.6 &                 &         &                   &         \\
357.866$-$0.997 &   881.6 &  15.17 & 357.865$-$0.995 &  $-$0.9 &                 &         &                   &         \\
357.886$+$0.004 &   292.4 &   6.62 & 357.885$+$0.006 &  $-$0.9 &                 &         & 357.885$+$0.005   &  $-$0.8 \\
357.907$+$0.107 &   129.0 &   5.71 & 357.907$+$0.109 &  $-$1.2 &                 &         & 357.906$+$0.109   &  $-$0.9 \\
358.003$-$0.637 &   654.5 &   6.62 & 358.003$-$0.636 &  $-$0.4 &                 &         &                   &         \\
358.118$+$0.006 &    83.2 &  10.38 &                 &         &                 &         & 358.116$+$0.007   &  $-$0.9 \\
\enddata

\footnote {Spectral index is defined as $S \sim \nu^{\alpha}$}

\end{deluxetable}

\begin{deluxetable}{llccc}
\tablewidth{0pc}
\tablecaption{Detected Pulsars \label{table:pulsars}}
\tablerefs{Taylor, Manchester, \& Lyne (1993) and the ATNF Pulsar Catalogue $\langle$http://www.atnf.csiro.au/research/pulsar/psrcat/$\rangle$}
\tablehead{

\colhead{Name}          &\colhead{Pulsar Name}            &\colhead{Angular Diameter} & 
\colhead{Flux Density}  &\colhead{$\alpha_{1.4}^{0.33}$}                             \\

& & \colhead{(\arcsec)} & \colhead{(mJy)}                 &

}

\startdata

 358.554$-$0.965 &  B1742$-$30   &   3.0 &     55.3 &   $-$0.5  \\
 359.305$-$0.841 &  J1747$-$2958 &   9.9 &     83.4 &   $-$0.7  \\
 359.460$-$1.246 &  B1746$-$30   &   1.6 &     24.0 &   $-$1.1  \\ 
   1.540$-$0.963 &  B1749$-$28   &   4.8 &   1099.3 &   $-$3.0  \\

\enddata

\end{deluxetable}

\begin{deluxetable}{lcccc}
\tablewidth{0pc}
\tablecaption{Pulsar non-detections\label{tab:missedpulsars}}
\tabletypesize{\small}
\tablehead{

\colhead{Pulsar Name}    & \colhead{Distance From}                 &
\colhead{Flux Density}   &\colhead{$3\sigma$ Flux Density\tablenotemark{a}}  &
\colhead{$\alpha_{1.4}^{0.33}$\tablenotemark{b}}\\

& \colhead{Phase Center(\arcmin)}   & \colhead{(mJy ($\nu=$1.4 GHz))}      & \colhead{(mJy ($\nu=$0.33 GHz))} & \colhead{(\mjybm)}

}

\startdata

J1738$-$2955 &  105  &    0.29    &     17.7   &$-$2.8\\      
B1736$-$29   &  80   &      2     &     9.9    &$-$1.1\\      
J1739$-$3023 &  114  &    1.0     &     21.3   &$-$2.1\\      
B1737$-$30   &  100  &    6.0     &     14.4   &$-$0.6\\      
J1740$-$3052 &  128  &    0.7     &     32.1   &$-$2.5\\      
J1741$-$2733 &  106  &    1.1     &     17.7   &$-$1.9\\      
J1741$-$2945 &  73   &    0.6     &     9.6    &$-$1.9\\      
J1741$-$3016 &  96   &    2.3     &     14.4   &$-$1.3\\      
J1747$-$2802 &  62   &    0.5     &     8.1    &$-$1.9\\      
J1752$-$2821 &  97   &    0.32    &     14.7   &$-$2.7\\

\enddata
\tablenotetext{a}{330 MHz flux density required for a $3\sigma$ detection at location of the pulsar.}
\tablenotetext{b}{Lower limit based on $3\sigma$ non-detection.}

\end{deluxetable}

\begin{deluxetable}{lccc}
\tablewidth{0pc}
\tablecaption{Pulsar Candidates \label{table:pcandidates}}
\tablehead{
 \colhead{Name} & \colhead{Angular Diameter} & 
\colhead{Flux Density} & 
   \colhead{$\alpha_{1.4}^{0.33}$} \\
  & \colhead{(\arcsec)} & \colhead{(mJy)}&}
\startdata

   357.907$+$0.107 &    5.71 &      129 &   $-$1.2 \\
   358.157$+$0.027 &    6.68 &    270.8 &   $-$1.0 \\
   358.556$-$0.572 &    5.03 &     40.3 &   $-$1.0 \\
   358.592$+$0.044 &    4.53 &    351.9 &   $-$1.0 \\
   358.638$-$1.162 &    5.77 &    251.7 &   $-$1.2 \\
   358.687$-$1.511 &     5.2 &     42.5 &   $-$1.5\tablenotemark{a} \\  
   358.756$+$0.972 &     7.7 &     27.6 &   $-$1.2\tablenotemark{a} \\  
   358.874$+$0.275 &     5.7 &    102.1 &   $-$1.9\tablenotemark{b} \\
\enddata
\tablenotetext{a} {Spectral index implied by non-detection in the Columbia Survey (Zoonematkermani et al. 1990 \& Helfand et al.~1992) and should be considered an upper limit.}
\tablenotetext{b} {Spectral index measured against the NVSS survey (Condon et al. 1998), not the Columbia survey.  The NVSS has a resolution of 45\arcsec, so spectral index values should be considered an upper limit.}
\end{deluxetable}

\clearpage

\begin{deluxetable}{lcccc}

\tablewidth{0pc}

\tablecaption{Previously Detected Non-Thermal Filaments }

\tablehead{

 \colhead{Name} & \colhead{Maximum Intensity}       & \colhead{Flux Density}   & \colhead {Size}      & \colhead {Plane Angle\tablenotemark{a}} \\

 \colhead{}    & \colhead{(mJy~beam${}^{-1}$)} & \colhead{(mJy)} & \colhead{(\arcmin)}  & \colhead {(\arcdeg)}}

\startdata

G0.08$+$0.15    &77.7   &4990   &$21.1 \times  0.3$   &0,10\tablenotemark{b} \\

G359.79$+$0.17  &48.0   &2540   &$8.3  \times  0.5$   &35\\

G359.54$+$0.18  &37.6   &1420   &$6.5  \times  0.8$   &25\\

G359.96$+$0.09  &25.7   &1450   &$12.5 \times  0.4$   &15\\

Radio Arc       &52.0   &24000  &$32.0 \times  4.0$   &5\\

Sgr C           &99.7   &4680   &$11.5  \times  0.2$   &15\\

G358.85$+$0.47  &7.7    &106    &$3.2  \times  0.8$   &90\\

G359.10$-$0.2   &22.4   &1540   &$22.7 \times  0.4$   &0, 10\tablenotemark{b} \\

G359.85$+$0.39  & \nodata & \nodata & \nodata & \nodata 

\enddata

\tablenotetext{a} {Plane Angle is the angle of the NTF with respect to the normal to the Galactic Plane.}

\tablenotetext{b} {This NTF shows significant 'kinks' and has therefore been fitted with two orientations.}

\label{table:oldntfs}

\end{deluxetable}

\begin{deluxetable}{lcccc}

\tablewidth{0pc}

\tablecaption{New Non-Thermal Filaments and Candidates }

\tablehead{

 \colhead{Name} & \colhead{Maximum Intensity}       & \colhead{Flux Density}   & \colhead {Size}      & \colhead {Plane Angle\tablenotemark{a}} \\

 \colhead{}    & \colhead{(mJy~beam${}^{-1}$)} & \colhead{(mJy)} & \colhead{(\arcmin)}  & \colhead {(\arcdeg)}}

\startdata

G359.12$+$0.66    &11.9  &647  &$15.6 \times  0.2$  &35\\

G359.22$-$0.16\tablenotemark{b}  &23.4  &269  &$1.8 \times   0.5$  &55\\

G359.33$-$0.42  &13.9  &81.0   &$2.4 \times   0.2$  &55\\

G359.36$+$0.09  &10.7  &65.2   &$2.5 \times   0.2$  &60\\

G359.40$-$0.03  &11.9  &93.8   &$1.8 \times   0.2$  &5\\

G359.40$-$0.07  &40.6  &229  &$1.7 \times   0.3$  &40\\

G359.43$+$0.13  &18.8  &265  &$2.4 \times   0.3$  &0,90\tablenotemark{c}\\

G359.59$-$0.34  &20.8  &188  &$2.3 \times   0.2$  &25\\

G359.66$-$0.11  &9.9   &226  &$3.5 \times   0.5$  &20\\

G359.85$-$0.02  &8.5   &173  &$1.8 \times   0.2$  &90\\

G359.86$-$0.24  &11.2  &205  &$8.1 \times   0.2$  &35\\

G359.88$-$0.07  &33.8  &930  &$1.6 \times   0.2$  &5\\

G359.90$+$0.19  &11.9  &129  &$2.4 \times   0.2$  &35\\

G359.99$-$0.54  &9.4   &88.8   &$8.6 \times   0.2$  &30\\

G0.02$+$0.04    &22.3  &228  &$2.0 \times   0.3$  &0\\

G0.06$-$0.07    &10.5  &163  &$2.1 \times   0.2$  &15\\

G0.37$-$0.07    &14.1  &128  &$1.1 \times   0.3$  &5\\

G0.39$+$0.05    &17.8  &232  &$4.1 \times   0.3$  &5\\

G0.39$-$0.12\tablenotemark{b}    &16.1  &731.2  &$10.1 \times  0.3$  &5\\

G0.43$+$0.01    &11.6  &43.9   &$1.6  \times  0.3$  &5\\

\enddata

\tablenotetext{a} {Plane Angle is the angle of the NTF with respect to the normal to the Galactic Plane.}

\tablenotetext{b}{Source observed to have significant 6 cm polarization (LaRosa et al.~2004.)}

\tablenotetext{c}{This source may be two interacting NTFs with orientations of 0\arcdeg\ and 90\arcdeg\ to the Galactic plane}

\label{table:newntfs}

\end{deluxetable}

\end{document}